\documentclass[]{jfm}

\usepackage{graphicx}
\usepackage{newtxtext}
\usepackage{newtxmath}
\usepackage{natbib}

\usepackage{hyperref}
\hypersetup{
    colorlinks = true,
    urlcolor   = blue,
    citecolor  = black,
}

\newcommand{\RomanNumeralCaps}[1]
\linenumbers

\usepackage{floatrow}
\usepackage[caption=false]{subfig}
\usepackage{threeparttable}
\usepackage[scientific-notation=true]{siunitx}
\usepackage{makecell}
\usepackage{commath}
\usepackage{chemformula}
\usepackage[scientific-notation=true]{siunitx}
\usepackage{booktabs}
\usepackage{cancel}

\usepackage{soul}

\title{A Kinetic Approach to Studying Low-Frequency Molecular Fluctuations in a One-Dimensional Shock}

\author{Saurabh S. Sawant\aff{1}
  \corresp{\email{sssawan2@illinois.edu}},
  D. A. Levin\aff{1}
 \and V. Theofilis\aff{2,3}}

\affiliation{\aff{1}Department of Aerospace, University of Illinois at Urbana-Champaign, 104 S. Wright St, Champaign, Illinois, USA
\aff{2} School of Engineering, University of Liverpool, The Quadrangle, Brownlow Hill, L69 3GH, UK
\aff{3} Escola Politecnica, Universidade S\~{a}o Paulo, Av. Prof. Mello Moraes 2231, CEP 5508-900, S\~{a}o Paulo-SP, Brasil
}

\begin{document}
\maketitle

\begin{abstract}
Low-frequency molecular fluctuations in the translational nonequilibrium zone of one-dimensional strong shock waves are characterised for the first time in a kinetic collisional framework in the Mach number range $2\le M\le 10$. 
Our analysis draws upon the well-known bimodal nature of the probability density function (PDF) of gas particles in the shock, as opposed to their Maxwellian distribution in the freestream, the latter exhibiting an order of magnitude higher dominant frequencies than the former.
Inside the (finite-thickness) shock region, the strong correlation between perturbations in the bimodal PDF and fluctuations in the normal stress suggests introducing a novel two-bin model to describe the reduced-order dynamics of a large number of collision interactions of gas particles. 
Our model correctly predicts the order-of-magnitude difference in fluctuation frequencies in the shock versus those in the freestream and is consistent with the small-amplitude fluctuations obtained from the highly resolved Direct Simulation Monte Carlo (DSMC) computations of the same configuration.
The variation of low-frequency fluctuations with changes in the conditions upstream of the shock revealed that these fluctuations can be described by a Strouhal number, based on the bulk velocity upstream of the shock and the shock-thickness based on the maximum density-gradient inside the shock, that remains practically independent of Mach number in the range examined.
Our results are expected to have far-reaching implications for boundary conditions employed in the vicinity of shocks in the framework of flow instability and laminar-turbulent transition studies of flows containing both unsteady and nominally stationary shocks.
\end{abstract}
\begin{keywords}
Authors should not enter keywords on the manuscript, as these must be chosen by the author during the online submission process and will then be added during the typesetting process (see \href{https://www.cambridge.org/core/journals/journal-of-fluid-mechanics/information/list-of-keywords}{Keyword PDF} for the full list).  Other classifications will be added at the same time.
\end{keywords}

\newpage

\section{Introduction}\label{Intro}
It is well-known that numerical solutions of the deterministic Euler and Navier-Stokes equations (NSE) require special treatment if one is to obtain reliable flow features containing shocks.
Solutions of differential equations describing fluid flow motion, in either a shock-capturing \citep[e.g.][]{lee1997interaction,priebe_tu_rowley_martin_2016} or a shock-fitting context \citep[e.g.][]{zhong1998high,sesterhenn2000characteristic} can predict both location and unsteadiness of shocks accurately, but the description of the internal shock structure and predictions of the gas properties inside a shock must be modelled in the context of such solutions.
Information on the interior of shocks may be delivered from the numerical solution of systems of equations generalising the Navier-Stokes equations.
The first documentation of differences in shock structure predicted by the Navier-Stokes equations and solution of the more general Bhatnagar-Gross-Krook model for the Boltzmann equation of transport is due to \citet{liepmann1962structure}, who showed that the Navier-Stokes equations are inadequate in describing the internal structure of a shock wave at high Mach numbers ($M\ge 3$) and that the differences between the two solutions increases in the low-pressure region of the shock layer with an increase in Mach number.
Soon afterwards, \citet{bird1970aspects} solved the exact Boltzmann equation using the stochastic DSMC method~\citep{bird:94mgd} and quantified the level of strong translational nonequilibrium in the interior of shocks.
\vspace{\baselineskip}

The significance of the correct description of not only the location and motion of shocks but also of their internal dynamical structure cannot be overstated in fluid mechanics. Amongst other classes of flows, shocks are responsible for the production of sound in supersonic jets as shown by the extensive studies of their interaction with turbulence and isolated vortices~\citep[e.g.][]{ribner1954a,ribner1954b,More1954,kovasznay1953turbulence,chang1957interaction,morkovin1962effects,mahesh1995interaction,mahesh1997influence,andreopoulos2000shock, larsson2009direct,koffi2008dynamics,xiao2014computational,singh2018non}.
The shock-wave/boundary-layer interactions (SBLIs) on compression ramps, cones, and flat plates are widely investigated for their role in generating separation bubbles, unsteadiness, and large surface heat and pressure fluxes~\citep[e.g.][]{dolling2001fifty,babinsky_harvey_2011,gaitonde2015progress}.
It is also recognised that the study of receptivity of shocks to freestream or induced disturbances is of importance in the investigation of transition in hypersonic boundary layers~\citep[e.g.][]{fedorov_2003, ma2003receptivity1, ma2003receptivity2, ma2005receptivity, hader2018towards}.
\vspace{\baselineskip}

Furthermore, research on the effect of kinetic fluctuations on triggering the transition from laminar to turbulent flows has been explored using Landau-Lifshitz's theory of fluctuating hydrodynamics (FH)~\citep{landauLifshitz} in several works on the receptivity of boundary layers in the incompressible~\citep{luchini2010thermodynamic,luchini2017receptivity} and high-speed compressible~\citep{fedorov2017receptivity, edwards2019model} regimes.
 In this phenomenological approach, stochastic fluxes, also called the Langevin source terms, are added to the stress tensor and heat flux vector in the NSE formulation to account for the effect of molecular fluctuations on the flow field.
The space-time correlation of these fluxes is given by Landau-Lifshitz's fluctuating-dissipation theorem in statistical mechanics for gas in equilibrium.
The modified governing equations are used to construct a receptivity problem, the solution of which gives the mean-square disturbance amplitude of macroscopic flow parameters excited by kinetic fluctuations.
Using this approach, \citet{fedorov2017receptivity} showed that in a compressible flat-plate boundary-layer, the kinetic fluctuations could trigger random wave-packets of Tollmien-Schlichting waves or Mack second-mode instability in the vicinity of lower neutral branch and their downstream growth could reach a threshold for the nonlinear breakdown.
However, this approach cannot be easily extended to account for the kinetic fluctuations in shocks, where the fluctuation amplitude is known to be larger than the predictions of the equilibrium theory~\citep[see][]{stefanov2000monte}.
Therefore, such investigations can benefit from kinetic methods that give detailed insights into the molecular origin of fluctuations in shocks, such as the specific nature and temporal changes of particles' velocity and energy distribution functions.
Understanding these details in two- and three-dimensional (2-D and 3-D) flows is challenging because of the additional effects of boundary-layers, instabilities, and unsteadiness of SBLIs.
Our analysis aims at closing this theoretical gap by examining the origin of molecular fluctuations in the well-known one-dimensional (1-D) shock of argon in a kinetic framework and show the surprising result that they exhibit \emph{low-frequency fluctuations}, that have been, up to the present, ignored in fluid mechanics literature. 
Their presence may well contribute to laminar to turbulent transition in supersonic and hypersonic flows.
\vspace{\baselineskip}

The internal structure of a normal shock has been a canonical case in numerous studies to understand thermal nonequilibrium in gases because of the absence of boundary-layer effects~\citep[e.g.][]{schmidt1969electron,alsmeyer1976density}.
Historically, solutions of the internal structure of strong shocks were sought using kinetic models~\citep{liepmann1962structure} as they allowed for anisotropy of stresses and heat fluxes, which are not accounted for by the traditional Navier-Stokes-Fourier constitutive relations.
\citet{bird1970aspects} successfully modelled the anisotropic shock structure using the DSMC method, which was also shown to agree well with the experiments of \citet{alsmeyer1976density}.
Since then, the method has been widely used for modeling 1-D shock structures~\citep[e.g.][]{cercignani1999structure, macrossan2003viscosity, ozawa2010particle, schwartzentruber2006hybrid, zhu2014modeling}.
Over the years, the method has also been shown to reproduce thermal fluctuations in larger systems of dilute gases~\citep[e.g.][]{garcia1986nonequilibrium, mansour1987fluctuating, garcia1991fluctuating, kadau2010atomistic, bruno2017rayleigh, bruno2019direct} and has been used successfully in simulating flow instabilities~\citep[e.g.][]{bird1998recent, stefanov_part1,stefanov_part2,stefanov_part3, kadau2004nanohydrodynamics, kadau2010atomistic, gallis2015direct,gallis2016direct}.
\vspace{\baselineskip}

Our past work has exploited the fidelity of DSMC to understand the critical role of shocks in hypersonic SBLIs.
\citet{tumuklu2018POF1,tumuklu2018POF2} simulated laminar SBLIs in a Mach 16 axisymmetric flow over a double-cone and observed strong coupling between a shock structure and a separation bubble.
For the freestream unit Reynolds number of $Re_1=\num{3.74e5}$~m$^{-1}$ they found oscillations of the detached (bow) and separation shocks characterised by a Strouhal number (nondimensional frequency) of 0.078.
In this work, we show that the Strouhal number associated with the low-frequency fluctuations in an isolated 1-D shock falls within a similar range of $St=$0.001 to 0.02.
\citet{sawant2018application} extended the capability of obtaining the DSMC solution on adaptively-refined octree grids and applied it to simulate challenging 3-D laminar SBLIs in a Mach 7 flow over a double-wedge at near-continuum input conditions corresponding to 59~km altitude.
Our recent works \citep{sawantIUTAM2019,sawant_SpanPeriodicDW} investigate the stability of the spanwise-periodic laminar separation bubble in the double-wedge flow to self-excited, small-amplitude, spanwise-homogeneous perturbations.
We elaborate on the coupling mechanism of the bubble and the shock structure and show, for the first time, that the instability of the bubble generates instability inside the strong gradient region of shocks.
As a result, the flow not only exhibits spanwise-periodic structures inside the separation bubble but also inside the separation and detached shocks, which results in spanwise modulation of shear layers downstream of triple points.
\vspace{\baselineskip}

With respect to macroscopic fluctuations in nonequilibrium zones,  the use of DSMC  has been a topic of extensive study in the context of imposed temperature gradients in a gas enclosed in isothermal walls~\citep[e.g.][]{garcia1986nonequilibrium, mansour1987fluctuating, ladiges2019suppression}; however, the literature on fluctuations in steady shock fronts characterised by extreme levels of nonequilibrium is sparse.
Notable of which is the work of \cite{stefanov2000monte}, who attributed an increase in velocity fluctuations in Mach 26 bow shock over a 2-D cylinder simulated by DSMC to `thermal nonequilibrium effects'; however, the study did not delve deeper into the reason for these effects, the changes in time-scales of fluctuations in comparison to an equilibrium state, and their dependence on the strength of the shock wave.
Our work attempts to answer all of these questions by investigating the fluctuations in molecular velocity and energy distribution functions obtained from the solution of the Boltzmann equation.
This work will show that the major role of bimodality in the distribution function inside the shock is to change the dominant frequencies of molecular fluctuations compared to  an equilibrium freestream.
\vspace{\baselineskip}

The paper is organised as follows:
Section~\ref{sec:Perturbations} describes the DSMC numerical setup and the nonequilibrium aspects of a 1-D shock structure of argon.
It then describes the fluctuations in the overall stress inside the shock with that in the freestream and points out the differences in their frequencies.
It is hypothesised that the differences in fluctuations are caused by the long-time collision interaction of particles in two modes of the bimodal energy distribution of particles.
To prove this hypothesis, in section~\ref{sec:2binModelcollis}, we construct a simplified two-energy-bin ordinary differential equation (ODE) model, similar to the predator-prey model of Lotka-Volterra~\citep{lotka1910,lotka1920,volterra1927fluctuations} but with modified terms accounting for intermolecular collisions between the two bins.
The evaluation of rate coefficients used in the model is also described in this section, whereas the simplification of the modelled collision processes is discussed in detail in the appendix~\ref{app:simplification}.
Section~\ref{sec:Dynamics} is devoted to the discussion of results obtained from the two-energy-bin model and their comparison with observations of fluctuations in the DSMC residuals.
Section~\ref{sec:Scaling} establishes a range of Strouhal numbers for Mach numbers 2 to 10 as well as for variations in upstream temperature at a given Mach number.
Finally, section~\ref{sec:Conclusion} summarizes the present findings.

\section{A Particle Representation of Nonequilibrium Fluctuations inside the Shock}~\label{sec:Perturbations}

\subsection{DSMC simulation methodology and properties of a 1-D shock}  
The DSMC solution  of the shock is obtained using the 1-D version of the Scalable Unstructured Gas-dynamic Adaptive mesh-Refinement (SUGAR) DSMC solver~\citep{sawant2018application}.  
A code-to-code validation of the DSMC solver was carried out with the numerical results of \citet{ohwada1993structure} in a 1-D, Mach three flow of argon.
Previously, Ohwada performed simulations using a finite-difference Boltzmann solver and obtained good agreement with the DSMC method for a hard sphere~\citep{bird:94mgd} collision model for this flow.
We obtained excellent agreement between our shock profiles of the normalized viscous stress and heat flux in the direction normal to the shock (not shown).
\vspace{\baselineskip}

The 1-D SUGAR code makes use of binary adaptive mesh refinement structure, where each computational Cartesian `root' cell is recursively refined into smaller cells until their size is smaller than the local mean-free-path.
The smallest cells, referred to as `leaf' or `collision' cells, are used to select neighbouring collision partners to perform binary elastic collisions between particles of a monatomic gas using the majorant frequency scheme~\citep{ivanov1988analysis}. 
The macroscopic flow and transport parameters are computed based on statistical equations of kinetic theory~\citep{bird:94mgd} and shown on larger root cells.
Therefore, each reference to a computational cell means a Cartesian root cell in this paper.
The gas is assumed to follow a variable hard sphere (VHS) molecular model~\citep[see][chapter 2, sec. 2.6]{bird:94mgd} with viscosity index $\omega=0.81$, mass $m=\num{6.637e-26}$~kg, reference diameter $d_{r}=\num{4.17e-10}$~m at reference temperature $T_{r}=273$~K.
\begin{figure}[H]
    \centering
    \sidesubfloat[]{\label{f:1D_Ar_Ttr}{\includegraphics[width=0.45\textwidth,height=0.40\textwidth]{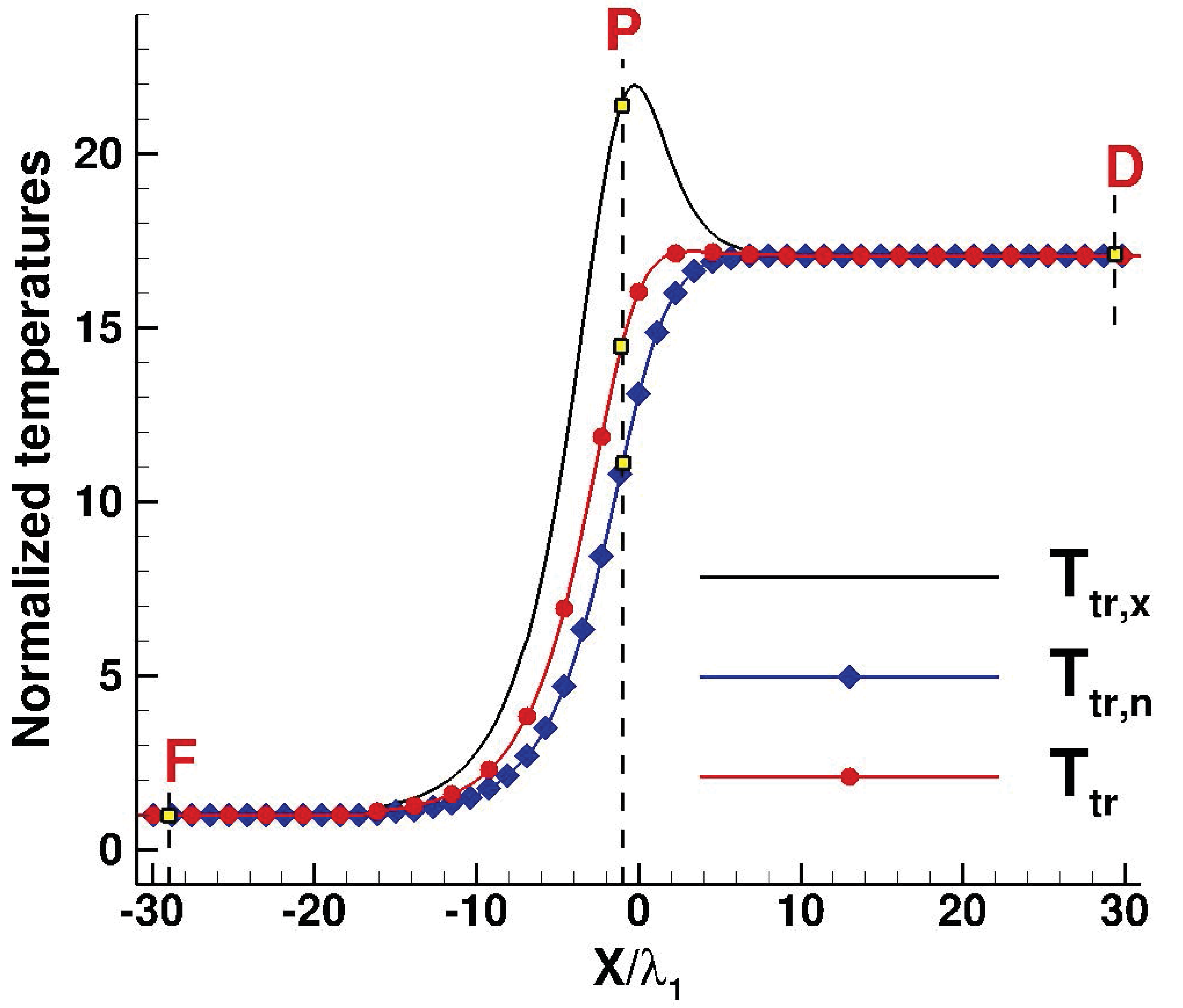}}}\hfill
    \sidesubfloat[]{\label{f:1D_Ar_pAndSigma}{\includegraphics[width=0.48\textwidth]{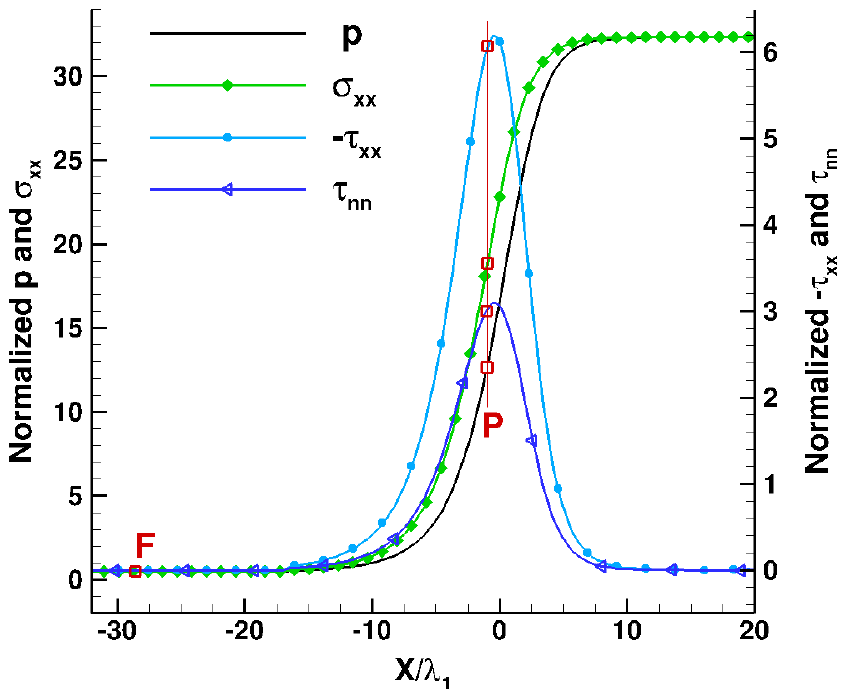}}}\,
    \sidesubfloat[]{\label{f:1D_Ar_uWithStd}{\includegraphics[width=0.47\textwidth]{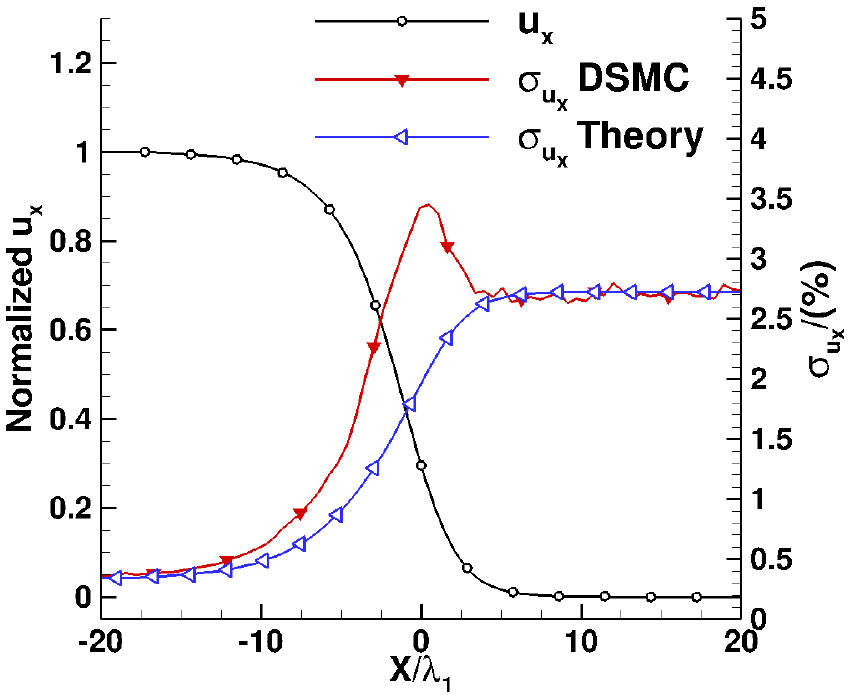}}}\hfill
    \sidesubfloat[]{\label{f:1D_Ar_nWithStd}{\includegraphics[width=0.47\textwidth]{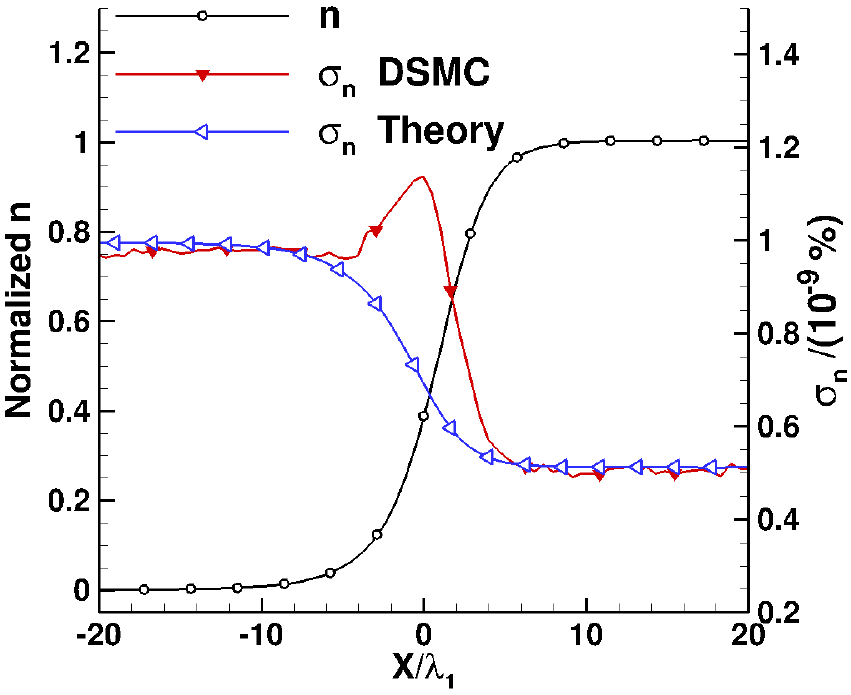}}}\,
    \caption{(\textit{a}) $T_{tr,x}$, $T_{tr,n}$, and $T_{tr}$ normalized by the upstream translational temperature $T_{tr,1}$. (\textit{b}) Normalized $p$, $\sigma_{xx}$, $\tau_{xx}$, and $\tau_{nn}$ (see text). (\textit{c}) and (\textit{d}) show normalized streamwise bulk velocity, $(u_{x}-u_{x,1})/(u_{x,2}-u_{x,1})$, and number density, $(n-n_1)/(n_2-n_1)$, respectively, as well as the comparison of DSMC-derived standard deviation in their fluctuations, $\sigma_{u_x}$ and $\sigma_{n}$, in percent of respective unnormalined macroscopic quantities with estimates from equilibrium statistical mechanics.}
\label{f:SampledTempAndTransport}
\end{figure}

The DSMC simulation was initialized from the Rankine-Hugoniot jump conditions at $X/\lambda_1=0$, where $X$ is the streamwise direction normal to the shock and $\lambda_1$ is the upstream mean-free-path.
The initial upstream flow conditions ($X/\lambda_1<$ 0) are hypersonic and downstream ($X/\lambda_1>$ 0), subsonic.
Subscripts `$1$' and `$2$' are used to denote the upstream and downstream macroscopic quantities.
The upstream boundary introduces directed local Maxwellian flow at a number density, bulk velocity, and temperature of $n_1=\num{1e22}$~m$^{-3}$, $u_{x,1}=3572.24$~m.s$^{-1}$, and $T_{tr,1}=710$~K, respectively, in the $X$-direction, whereas the downstream boundary mimicks a specularly reflecting solid surface that moves at a bulk velocity of $u_{x,2}$, equal to the downstream Rankine-Hugoniot velocity, as described by \citet[see chapter 12]{bird:94mgd}.
The flow takes approximately 6~$\mu$s to transition from the jump condition to the steady state shock structure.
Particles have zero lateral bulk velocities but non-zero thermal velocities in all three directions.
Bird's STABIL boundary condition~\citep[see chapter 12]{bird:94mgd} is also used to prevent random walk effects from causing the shock to move; however, because of the use of a large number of computational particles ($\approx$475,000), it would take much longer time ($>$ 0.45~ms) well beyond the relevant time-scales of interest in this work for such numerical effects to become important.
\vspace{\baselineskip}

With this approach, figure~\ref{f:SampledTempAndTransport} shows time-averaged profiles of macroscopic flow and transport parameters obtained from the DSMC simulation of a 1-D, Mach 7.2 shock in argon.  
The time-averaged or mean quantities mentioned in this work were calculated for 30~$\mu$s after the transient period of 6~$\mu$s, and no difference was found if the results were time-averaged for a longer period of 0.18~ms. 
Figure~\ref{f:1D_Ar_Ttr} shows deviation of the $X-$ and average lateral directional temperatures, $T_{tr,x}$ and $T_{tr,n}=(T_{tr,y}+T_{tr,z})/2$, respectively, from the overall translational temperature, $T_{tr}$, within a region of $-16 < X/\lambda_1 < 6$, indicating the presence of translational nonequilibrium in the shock.
$T_{tr,n}$ is averaged owing to the axial symmetry between $Y$ and $Z$-directions.
The directional temperatures are a measure of average thermal energies of particles in Cartesian directions~\citep[see][sec. 1.4]{bird:94mgd}, whereas the overall translational temperature is obtained by averaging the former.
In the equilibrium region ($X/\lambda_1 < -16$ and $X/\lambda_1 > 6$), all directional temperatures are equal to the overall temperature.
Within the region $2< X/\lambda_1 < 6$ the gradient of overall translational temperature reaches zero, yet there is a strong deviation of directional temperatures from each other, as was observed by the DSMC simulation of a Mach 8 argon shock by \citet{bird1970aspects}.
\vspace{\baselineskip}

Additionally, figure~\ref{f:1D_Ar_pAndSigma} shows the deviation of time-averaged pressure, $p$, from the $X$-directional overall stress, $\sigma_{xx}$, inside the shock as well as the viscous stress components, $\tau_{xx}$ and $\tau_{nn}=(\tau_{yy}+\tau_{zz})/2$, calculated based on the relationship,
\begin{equation} 
\begin{split}
\tau_{ij}&=-(\sigma_{ij} - p\delta_{ij}) 
\end{split}
\label{NS_1D_ViscousStress}
\end{equation}
where $\delta_{ij}$ is the Kronecker delta function.
The pressure and stress components are normalized by the upstream parameter $\rho_1\beta_1^{-2}$, where $\beta=\sqrt{m/2\kappa_b T_{tr}}$ is the inverse of the most probable speed of molecules, $m$ is mass, $\kappa_b$ is the Boltzmann constant, $\rho=nm$ is the mass density, and $n$ is the number density. The stress components, $\tau_{ij}$ and $\sigma_{ij}$, act in the $i^{th}$-direction on a plane with normal in the $j^{th}$-direction.
The existence of these non-zero stresses leads to a finite thickness of the shock wave, which, when modelled by accounting for molecular thermal fluctuations using a stochastic method, such as DSMC, reveals nonequilibrium bimodal velocity and energy distributions of particles.
This paper will show that these bimodal distributions exhibit an order of magnitude lower dominant frequency fluctuations than those found in the freestream.
\vspace{\baselineskip}

With respect to the equilibrium regions of the shock, figures~\ref{f:1D_Ar_uWithStd} and~\ref{f:1D_Ar_nWithStd} show excellent agreement between the DSMC-computed velocity and number density fluctuations and theory, as was observed by  \cite{stefanov2000monte}.  Theory predicts that at equilibrium, statistical fluctuations in macroscopic quantities of bulk velocity, $u_x$, and number of particles, $N$, have a standard deviation of $\sqrt{\kappa_b\langle T_{tr}\rangle/m\langle N\rangle}$ and $\sqrt{\langle N \rangle}$, respectively, where the latter is a simplified result for a dilute gas~\citep[][chapter XII]{hadjiconstantinou2003statistical, landauLifshitz}. 
The brackets denote time or ensemble average of macroscopic quantities.
The DSMC standard deviations are computed for each computational cell of width $\Delta x=\num{1e-4}$~m for instantaneous data collected at every timestep of $\Delta t=3$~ns from 6~$\mu$s to 0.2~ms. Note that since the velocity fluctuations shown in figure~\ref{f:1D_Ar_uWithStd} are expressed in terms of DSMC computational particles, the true amplitude of the  actual thermal fluctuations is obtained by multiplication of $\sqrt{FNUM}$, where $FNUM=10^7$ is the number of dilute gas molecules represented by each DSMC computational particle~\citep{bruno2019direct, hadjiconstantinou2003statistical, stefanov_part2, bruno2019direct}.   Therefore, 
the 0.344~\% standard deviation of velocity fluctuations in the freestream in figure~\ref{f:1D_Ar_uWithStd} corresponds to an actual value of $\num{1.09e-4}$~\%,   a small yet significantly large number on the scale of small amplitude perturbations considered in shock-dominated flows.
A minor point to note is that the velocity fluctuations increase downstream due to the increase in mean temperature and decrease in bulk velocity which dominate over the increase in the mean number of particles.

\subsection{Nonequilibrium fluctuations} 
We are particularly interested in the nonequilibrium zone of the shock layer, where a significant deviation from equilibrium is seen in DSMC-computed standard deviations, as shown in figures~\ref{f:1D_Ar_uWithStd} and~\ref{f:1D_Ar_nWithStd}.
Note that the standard deviations peak at the location of maximum gradients ($X/\lambda_1=0$) of the respective flow parameters.
In contrary to the findings of \cite{stefanov2000monte}, the density fluctuations inside a shock also deviate from the equilibrium Poisson law, although their magnitude is much smaller than the velocity or energy fluctuations.
\begin{figure}[H]
    \centering
    \sidesubfloat[]{{\label{f:ProbeDataStresses_a}\includegraphics[width=0.49\textwidth]{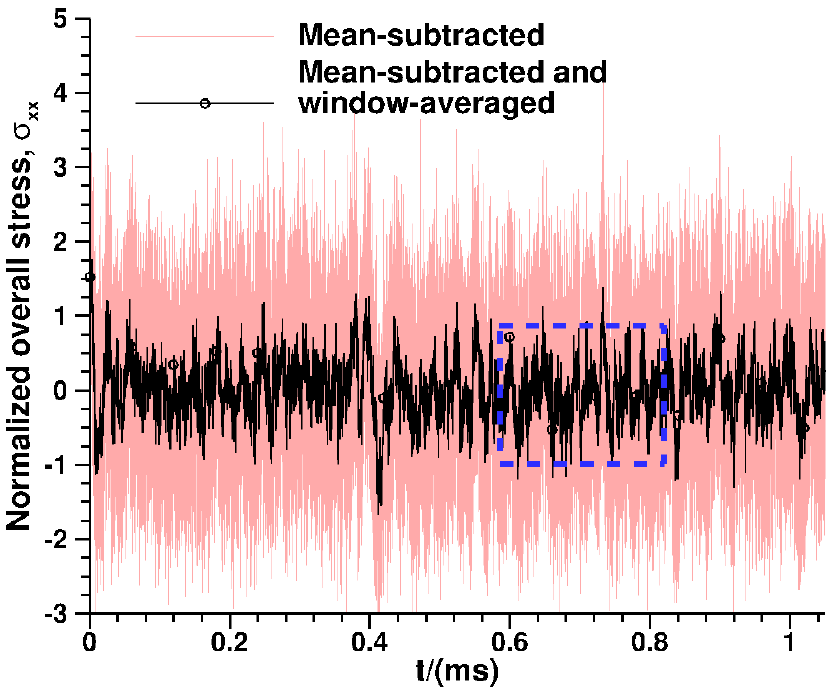}}}\hfill
    \sidesubfloat[]{{\label{f:ProbeDataStresses_b}\includegraphics[width=0.45\textwidth]{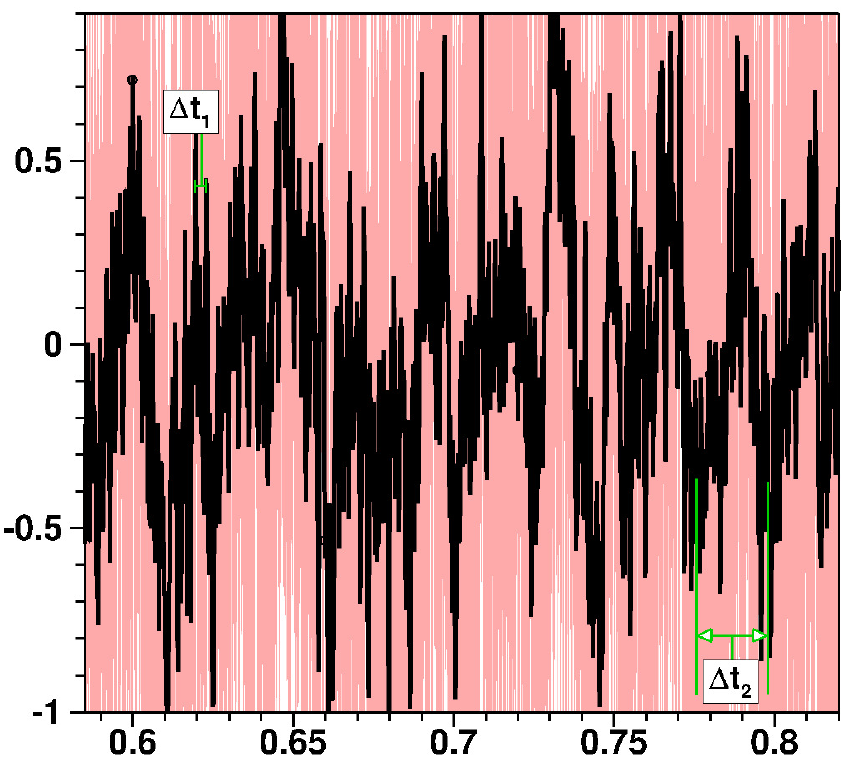}}}\,
    \caption{(\textit{a}) The instantaneous mean-subtracted and then window-averaged data of $\sigma_{xx}\rho_1^{-1}\beta_1^{2}$ at numerical probe $P$ located at $X/\lambda_1=-1$, as indicated in figure~\ref{f:1D_Ar_pAndSigma}. (\textit{b}) Zoom of the region marked by a dashed box in (\textit{a}). Note that $\Delta t_1^{-1}$=286~kHz and $\Delta t_2^{-1}$=46~kHz. The window-average is obtained by taking a moving average of the mean-subtracted instantaneous data at every 0.3~$\mu$s (100 timesteps) such that frequencies greater than 3.33~MHz are filtered off.}
\label{f:ProbeDataStresses}
\end{figure}

\begin{figure}[H]
  \sidesubfloat[]{\label{f:PSD_ProbeP}{\includegraphics[width=0.46\textwidth]{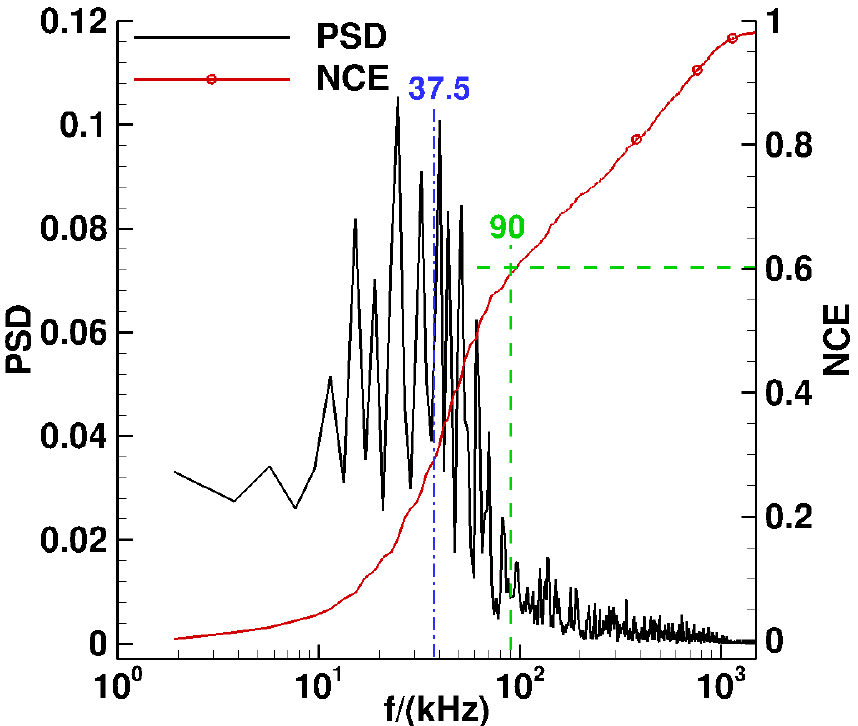}}}\hfill
    \sidesubfloat[]{\label{f:PSD_ProbeF}{\includegraphics[width=0.46\textwidth]{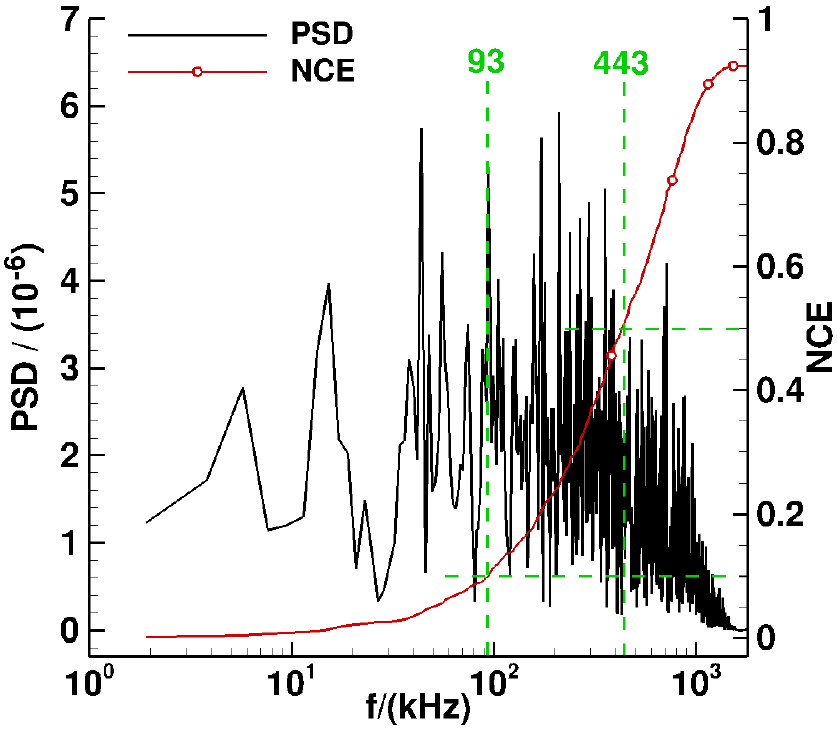}}}\,
    \sidesubfloat[]{\label{f:PSD_Contour}{\includegraphics[width=0.55\textwidth]{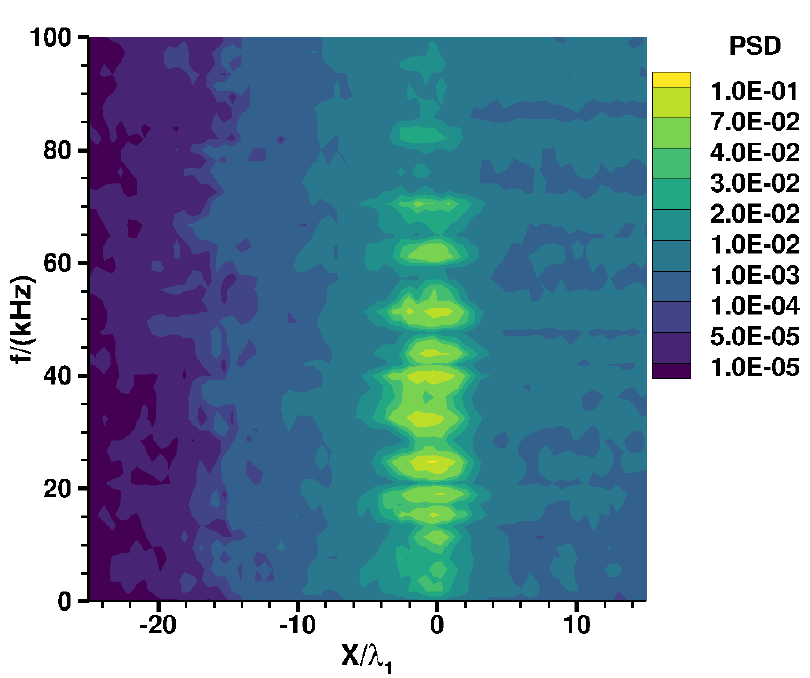}}}\,
    \caption{The PSD obtained from the mean-subtracted, window-averaged data of $\sigma_{xx}\rho_1^{-1}\beta_1^{2}$ at (\textit{a}) probe $P$ located at $X/\lambda_1=-1$ and (\textit{b}) probe $F$ located at $X/\lambda_1=-29$ along with the NCE (see text). (\textit{c}) Contours of PSD along the entire $X/\lambda_1$.}
\label{f:PSD_SigmaXX}
\end{figure}

Figure~\ref{f:ProbeDataStresses_a} shows at probe $P$ ($X/\lambda_1=-1$), the time history of fluctuations about the time-averaged mean value of normalized $X$-directional overall stress, $\sigma_{xx}\beta_1^2\rho_1^{-1}$, which can also be written as $\rho T_{tr,x} (2T_{tr,1})^{-1}$ based on the definition of $\beta_1$ and $\sigma_{xx}=\rho R T_{tr,x}$.
Since the density fluctuations are negligible, fluctuations in $\sigma_{xx}$ correspond to those in $T_{tr,x}$.
To observe the low-frequency fluctuations with more clarity, another signal is overlaid, which is obtained by window-averaging the instantaneous signal with a moving time-window of 0.3~$\mu$s (100 timesteps).
It reveals two disparate frequencies of 286 and 46~kHz, as shown in figure~\ref{f:ProbeDataStresses_b}.
Similar frequencies are observed in other macroscopic flow parameters such as other directional temperatures, viscous stresses, pressure, and velocities (not shown).
The power spectral density (PSD) of the mean-subtracted, window-averaged data of $\sigma_{xx}$ is shown in figure~\ref{f:PSD_SigmaXX}.
For spectral estimation here and elsewhere in the paper, Welch's method~\citep{welch1967use,solomon1991psd} is used in \cite{SciPy} software with two Hann-window weighted segments of data sampled with a frequency of 333~MHz prior to the Fast Fourier Transform (FFT) such that the frequency resolution is 0.9~kHz.
At probe $P$, the PSD shows a broadband of low-frequencies that ranges up to 90~kHz frequency, as shown in figure~\ref{f:PSD_ProbeP}.
This upper bound of the broadband is defined as the frequency at which the normalized cumulative energy (NCE), obtained from normalizing and cumulatively summing the PSD spectrum, exhibits an inflection point. 
The broadband contains nearly 60\% of the total spectral energy and can be characterised by its weighted average of 37.5~kHz with a standard deviation of 21.4~kHz.
Furthermore, figure~\ref{f:PSD_Contour} shows the contours of PSD plotted on the axes $X/\lambda_1$ versus frequency and reveals that such low-frequency broadband is expected within $-6 < X/\lambda_1 < 3$, a region of strong translational nonequilibrium.
Note that the contours are created by interpolating the PSD data at each $X/\lambda_1$ location spatially separated by $X/\lambda_1=1$ in the \citet{Tecplot} software using the inverse-distance algorithm with default parameters (exponent=3.5, point selection=Octant, Number of points=8). 

In comparison, the spectrum at probe $F$ in the freestream, shown in figure~\ref{f:PSD_ProbeF}, contains widely distributed energy across the whole spectral limit of 1666~kHz and does not exhibit a noticeable inflection point within this limit.
Note, however, that 40\% of the total spectral energy and many peaks are located within a band of 93 to 443~kHz, which correspond to fluctuations with an order of magnitude higher time-scales than the mean collision time of 0.284~$\mu$s.
This can be compared with the gas in absolute equilibrium (zero bulk velocity, constant temperature and number of particles), where small perturbations to velocity distribution function and its moments decay exponentially with a characteristic time-scale of mean collision time~\citep[see][pg. 82, 124]{koganRGDSpringer}.
However, the prior statement is based on the analysis which assumes equal relaxation time for all molecules at a given space and time in the entire velocity space.
More importantly, an order of magnitude differences in time-scales of fluctuations in the freestream versus the shock are explained in section~\ref{sec:Dynamics}.
\begin{figure}[H]
    \centering
    \sidesubfloat[]{\label{f:fex_F}{\includegraphics[width=0.3\textwidth]{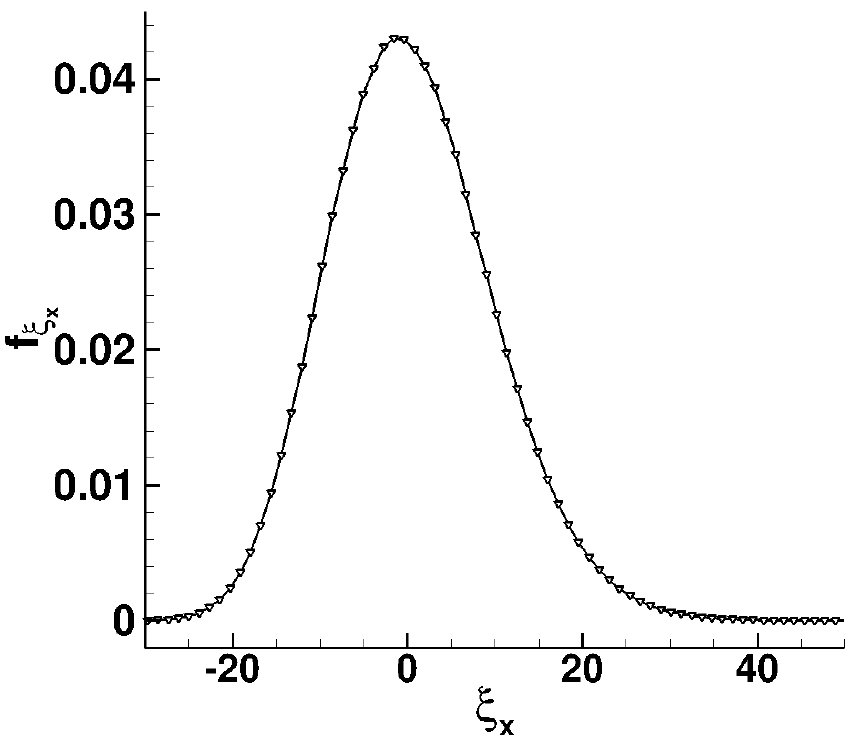}}}\hfill
    \sidesubfloat[]{\label{f:fex_D}{\includegraphics[width=0.3\textwidth]{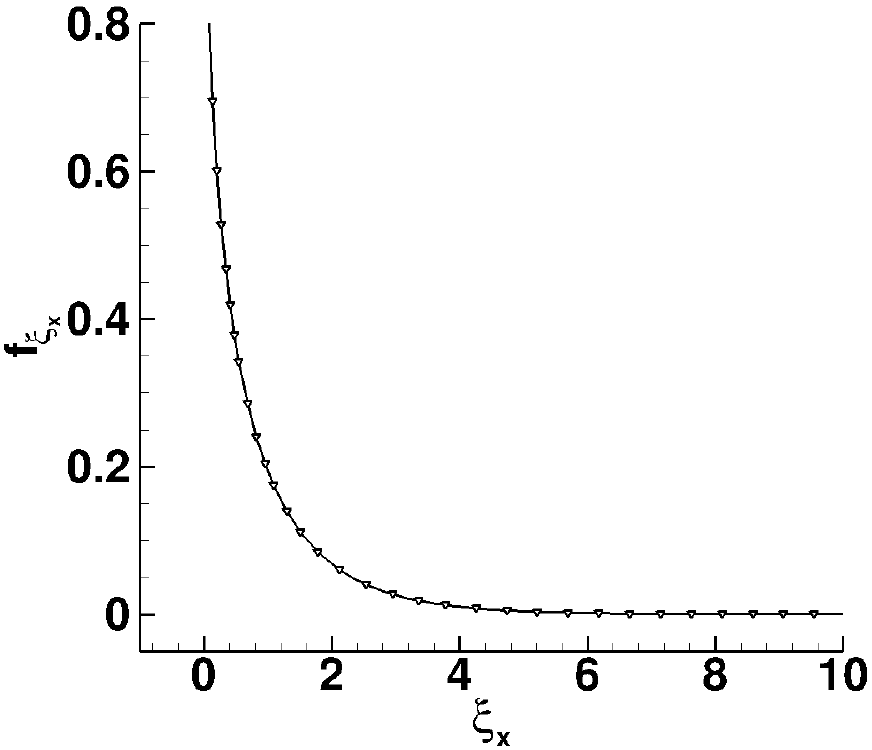}}}\hfill
    \sidesubfloat[]{\label{f:fex_P}{\includegraphics[width=0.3\textwidth]{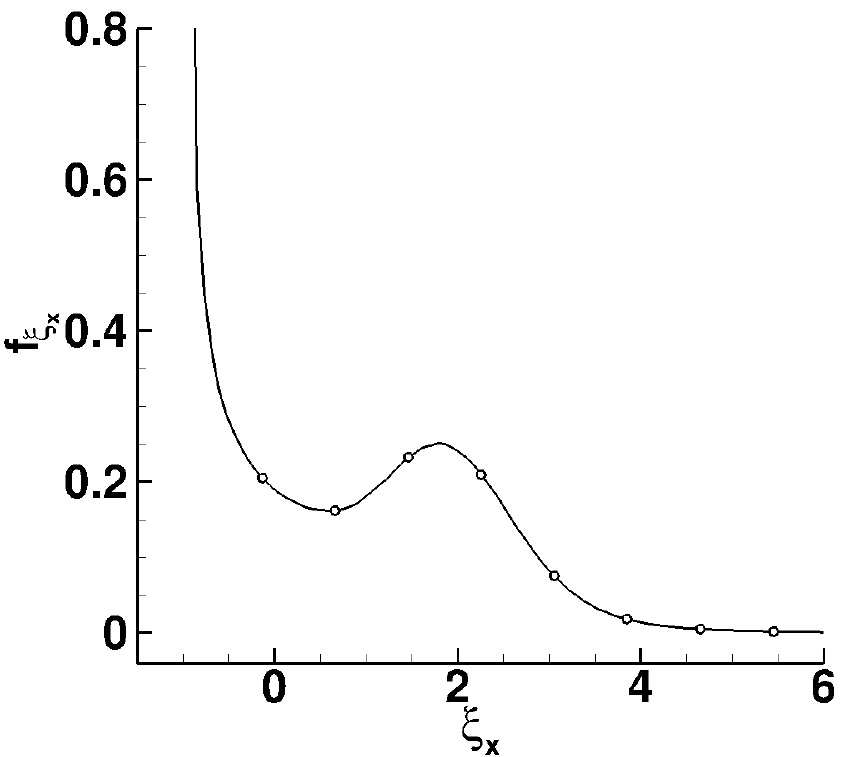}}}\hfill
    \caption{The PDF $f_{\xi_x}$ at probes (\textit{a}) $F$ in the freestream ($X/\lambda_1=-29$), (\textit{b}) $D$ downstream ($X/\lambda_1=29$), and (\textit{c}) $P$ inside the shock layer ($X/\lambda_1=-1$).}
    \label{f:fex}
\end{figure}

The key to understanding the fluctuations in the overall stress, $\sigma_{xx}$, is to correlate the fluctuations in the PDF $f_{\xi_x}$ of the normalized $X$-directional energy of particles, $\xi_x$, and its mean, $\mu$. 
Note that $\xi_x=(v_x^2-u_x^2)\beta^2$, where $v_x$ is the $X$-directional instantaneous molecular velocity of particles, which is the sum of their bulk and thermal components, $u_x$ and $c_x$, respectively.
The mean, $\mu$, is related to overall stress, $\sigma_{xx}$, as,
\begin{equation} 
\begin{split}
\mu = \int{\xi_x f_{\xi_x}} d\xi_x &= \frac{\sigma_{xx}\beta^2}{\rho}\\
\end{split}
\nonumber
\end{equation}
Figure~\ref{f:fex} shows that the behavior of the PDF changes at different locations in the flow, {\em i.e.,} from the upstream to downstream regions relative to the shock,  $f_{\xi_x}$ changes from a nearly symmetric equilibrium distribution (figure~\ref{f:fex_F}) to a one-sided, asymmetric equilibrium distribution (figure~\ref{f:fex_D}). At  the location of maximum density gradient in the shock (figure~\ref{f:fex_P}), it can be seen that the density function is bimodal with an inflection point at $\xi_x=1.33$.
Note that the bimodal PDF can be expressed as a linear combination of upstream and downstream contributions of equilibrium PDFs, similar to the Mott-Smith model of the bimodal velocity distribution.
\vspace{\baselineskip}

Towards that end, the cross-correlation coefficient of $\mu$ and the number of particles as a function of $\xi_x$ is defined as,
\begin{equation} 
\label{CCF}
\centering
c(\xi_x) = \frac{\sum_{w=0}^{w=W} \sum_{\xi_x=min}^{max} {\left[N(\xi_x,w)-\langle N(\xi_x) \rangle_{w}\right]\left[\mu(w)-\langle\mu\rangle_w\right]}}{W \Sigma_{N(\xi_x)} \Sigma_{\mu_{\xi_x}}}
\end{equation}
where
\begin{eqnarray}
\notag
\Sigma_{N(\xi_x)} &=& \sqrt{\frac{\left[N(\xi_x,w)-\langle N(\xi_x) \rangle_{w}\right]^2}{W}}\\
\notag
\Sigma_{\mu} &=& \sqrt{\frac{\left[\mu(w)-\langle\mu\rangle_w\right]^2}{W}}
\end{eqnarray}
Note that $\xi_x$ is discretized into 200 energy bins from its minimum to maximum value.
$N(\xi_x,w)$ is the total number of  DSMC particles within the normalized $X$-directional energy space $\xi_x$ and $\xi_x+\Delta\xi_x$ from time window $w$ to $w+1$,  $\langle N(\xi_x) \rangle_{w}$ denotes the number of particles in the same energy space but averaged over  time-windows, $W$.
Similarly, $\mu(w)$ is the instantaneous mean of the PDF $f_{\xi_x}$, from time-window $w$ to $w+1$, whereas $\langle\mu\rangle_w$ is the mean computed by averaging over all time-windows.
$\Sigma_{N(\xi_x)}$ and $\Sigma_{\mu}$ are the standard deviations in the fluctuations of $N(\xi_x,w)$ and $\mu(w)$ about their respective means.  
Supplementary movies 1 and 2 show  the fluctuation $\left[N(\xi_x,w)-\langle N(\xi_x)\rangle_w\right]$ as a function of $\xi_x$ at probes $P$ and $F$, respectively.\footnote{The movies loop over time windows from -20 to 646, where $w=$-20 to 0 correspond to the transient time, which is not used in the calculation of cross-correlation coefficient and time-averaged means.}
\begin{figure}[H]
    \centering
    \sidesubfloat[]{\label{f:CCF_ProbeP}{\includegraphics[width=0.44\textwidth]{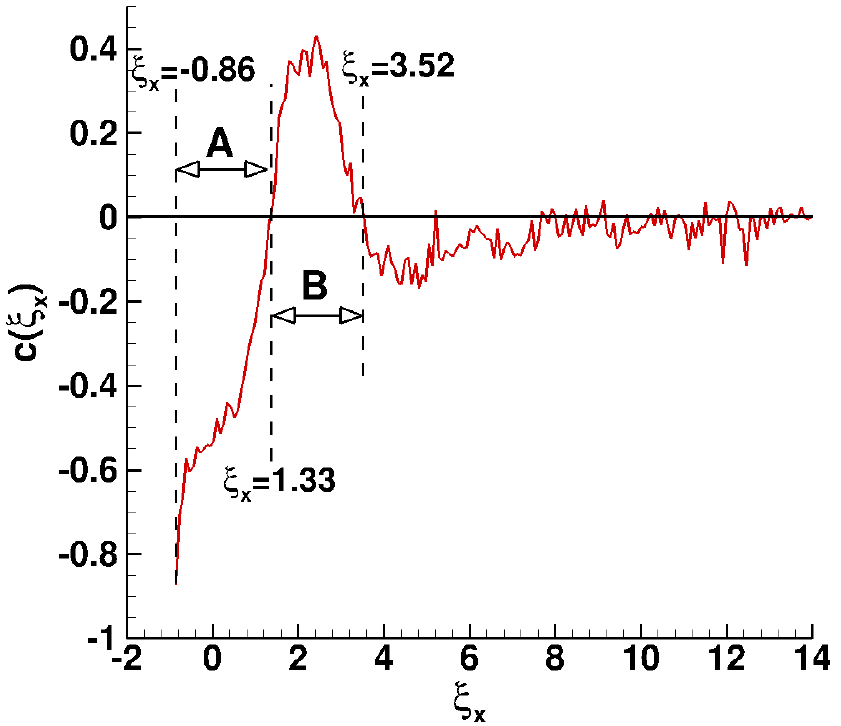}}}\hfill
    \sidesubfloat[]{\label{f:CCF_ProbeF}{\includegraphics[width=0.44\textwidth]{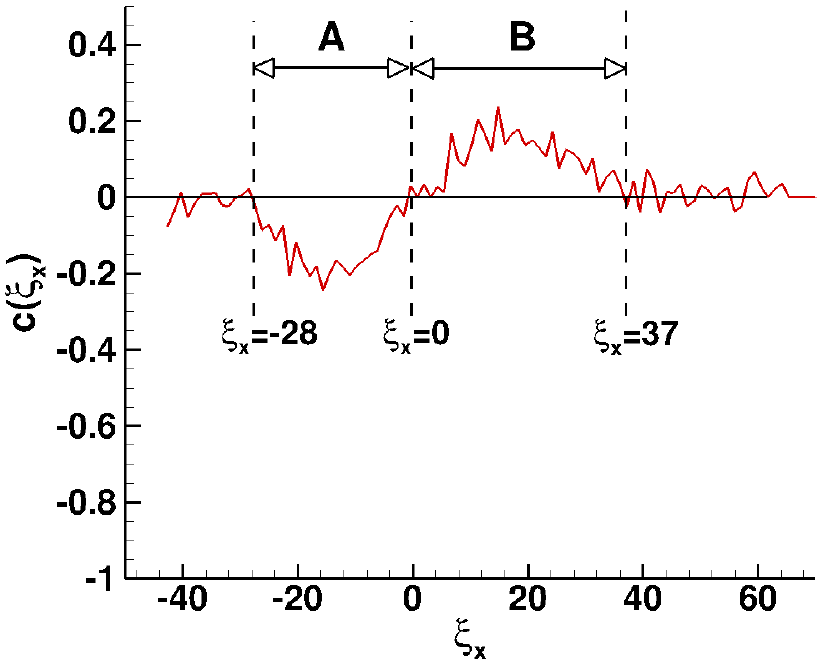}}}\,
    \sidesubfloat[]{\label{f:NaNb_ProbeP}{\includegraphics[width=0.44\textwidth]{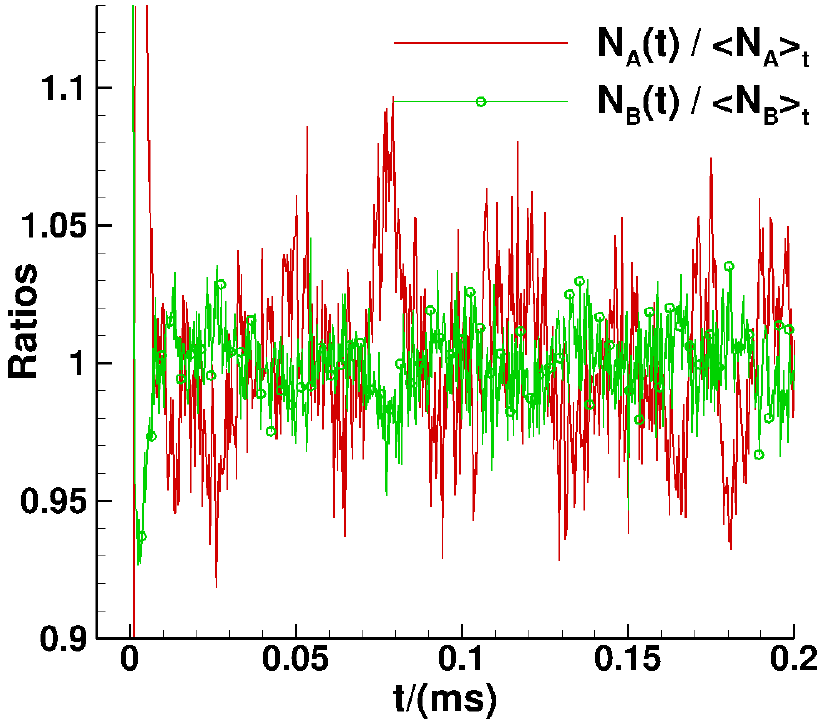}}}\hfill
    \sidesubfloat[]{\label{f:NaNb_ProbeF}{\includegraphics[width=0.44\textwidth]{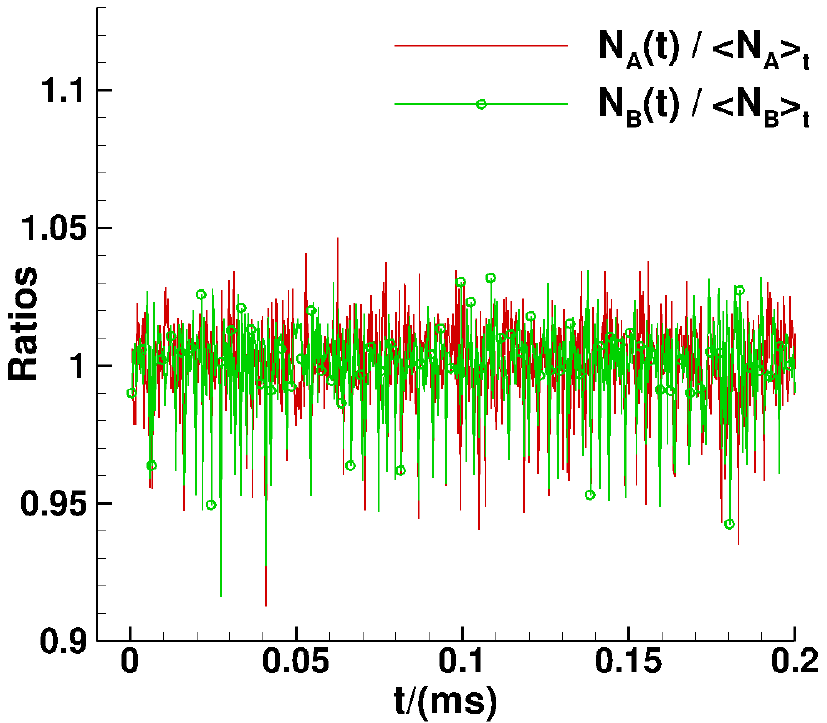}}}\,
    \caption{(\textit{a}) and (\textit{b}) show $c(\xi_x)$ calculated using equation~\ref{CCF}, at probes $P$ and $F$, respectively.    Here and in subsequent figures, the bracket notation, $\langle Q\rangle_t$ denotes the time-average of a macroscopic quantity $Q$, in contrast to its instantaneous value. 
     (\textit{c}) and (\textit{d}) show the ratios of instantaneous to time-averaged number of particles in bins $A$ and $B$, i.e., $N_A/\langle N_A \rangle_t$ and $N_B/\langle N_B \rangle_t$, at probes $P$ and $F$, respectively. 
 To reduce statistical scatter, the instantaneous number of particles is averaged over a very small time internal of 0.3~$\mu$s, 
 whereas the time-averaged number of particles is averaged from 6~$\mu$s to 0.2~ms.}
    \label{f:energyBinsAandB}
\end{figure}

The analysis of the correlation functions combined with particle distributions suggests an approach for grouping of particles into a finite number of energy bins whose dynamics can then be further analysed.  
Starting with figure~\ref{f:energyBinsAandB}, the distribution of $c(\xi_x)$ at probes $P$, inside a shock, and $F$, in the freestream are compared.  
$c(\xi_x)$ is calculated using a total number of time windows, $W$, of 646, each representing a time interval of 0.3~$\mu$s starting from $t$=6~$\mu$s.
At probe $P$, figure~\ref{f:CCF_ProbeP} shows that $c(\xi_x)$ has a strong negative correlation 
 between $-0.86 < \xi_x < 1.33$, becomes positive for $1.33 < \xi_x < 3.52$, and remains small but negative for $\xi_x > 3.52$.  
The demarcation between the coarse energy bins $A$ and $B$ is defined as $c(\xi_x)=0$, which is also the location of inflection point.
Examination of $c(\xi_x)$ for probe $F$ in the freestream  (see figure~\ref{f:CCF_ProbeF}) shows that, in contrast, the maximum magnitude of $c(\xi_x)$ is not more than 0.2, indicating that the fluctuations are random in nature and even if a strong correlation exists, it must be on time-scales smaller than the window size of $0.3$~$\mu$s used to obtain $c(\xi_x)$.  Comparison of the correlation functions at probes $P$ and $F$ also shows that   the distribution of $c(\xi_x)$ at the latter location is almost symmetric about the inflection point at $\xi_x=0$, which is consistent with the 
 $X$-directional energy distribution having a nearly symmetric shape, as shown in figure~\ref{f:fex_F}. 
\vspace{\baselineskip}

If we now group particles into two $X$-directional energy bins, `$A$' and `$B$', we can study the ratio of the instantaneous to time averaged particles, $N_A/\langle N_A \rangle_t$ and $N_B/\langle N_B \rangle_t$ as a function of time, as shown in figures~\ref{f:NaNb_ProbeP} with~\ref{f:NaNb_ProbeF}.  
The average number of DSMC computational particles per computational cell  in energy bins $A$ and $B$ are $\langle N_A \rangle_t=1088$ and $\langle N_B \rangle_t=611$ at probe $P$, and $\langle N_A \rangle_t=510$ and $\langle N_B \rangle_t=493$ at probe $F$, which can be converted to  
number density $(m^{-3})$ by multiplication of an FNUM=$\num{1e7}$ and division by the volume of computational cell, $\Delta v=\num{1e-12}~m^3$.
From the comparison of these two figures, it can be seen that fluctuations in this ratio are larger 
 in magnitude and exhibit longer time-scales at probe $P$ inside the shock than at probe $F$ in the freestream.
The standard deviations in the fluctuations of the ratios $N_A/\langle N_A \rangle_t$ and $N_B/\langle N_B \rangle_t$ are 4.5 and 2.1\%, respectively, at probe $P$, versus 1.8 and 1.6\% at probe $F$.
Additionally, a noticeable negative correlation between the fluctuations of two energy bins at probe $P$ can also be seen from figure~\ref{f:NaNb_ProbeP}, while such dynamics are absent at probe $F$.

\begin{figure}[H]
    \centering
    \sidesubfloat[]{\label{f:AvgParticles_ProbeP}{\includegraphics[width=0.46\textwidth]{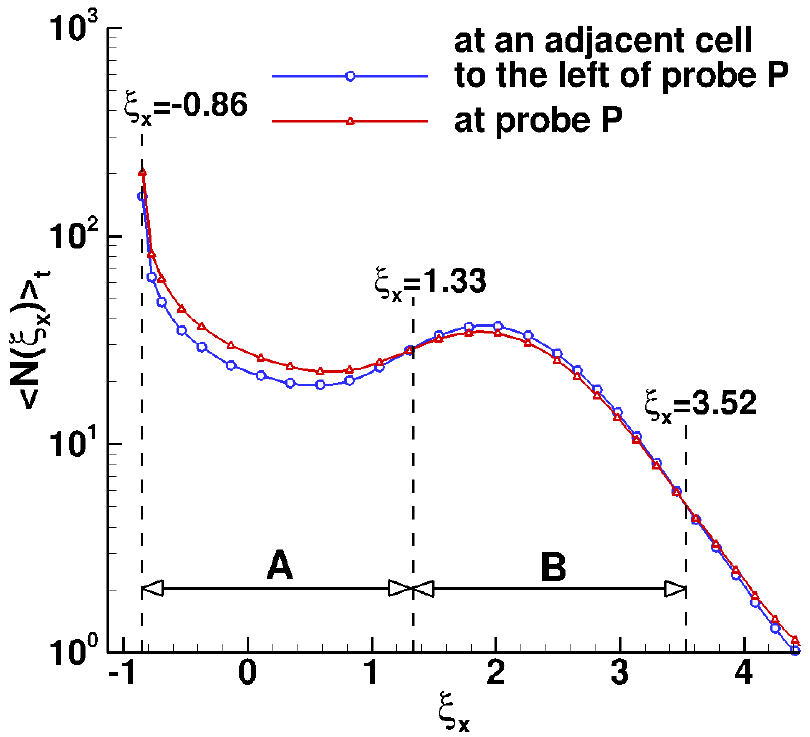}}}\hfill
    \sidesubfloat[]{\label{f:AvgParticles_ProbeF}{\includegraphics[width=0.46\textwidth]{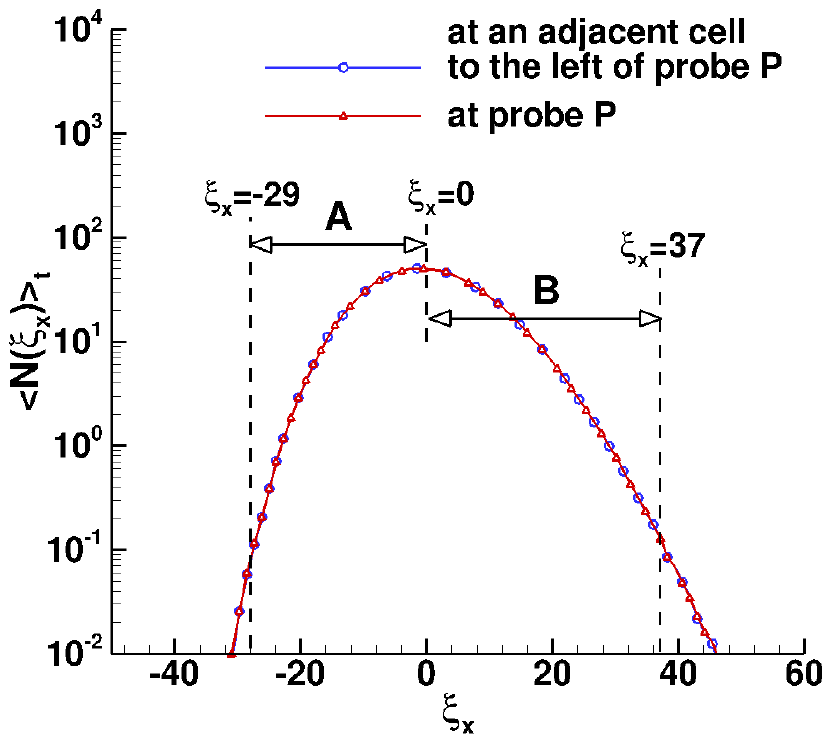}}}\hfill
    \caption{(\textit{a}) and (\textit{b}) show a comparison of the average number particles as a function of $X$-directional energy, $\langle N(\xi_x) \rangle_t$, at probes $P$ and $F$, respectively, with those in an adjacent cell to their left. The time-averages are taken from t=6 $\mu$s to 0.2~ms.}
    \label{f:AvgParticles}
\end{figure}
Finally, to demonstrate the generality of the difference in properties between probes $P$ and $F$, 
figure~\ref{f:AvgParticles}  shows a comparison of the time-averaged number of particles as a function of $\xi_x$ at probes $P$ and $F$, and at their respective left-adjacent computational cells.
For probe $P$ (figure~\ref{f:AvgParticles_ProbeP}), two inflection points can be identified at $\xi_x=1.33$ and $3.52$, where $c(\xi_x)$ changes sign.
We can also see that the average number of particles in energy bin $A$ is larger at probe $P$ than in the cell to its left, whereas in bin $B$ it is lower.
This is consistent with the fact that from the upstream to downstream region inside a shock, the contribution of the upstream symmetric distribution decreases, while the downstream asymmetric distribution increases. 
At probe $F$ (figure~\ref{f:AvgParticles_ProbeF}), however, the PDF of energies does not change between adjacent cells and there is no inflection point.  

\section{Collision processes that define the two-bin model}~\label{sec:2binModelcollis}

Although the distribution of $X$-directional particle energies is continuous, we propose a two-bin model to understand the role of particle collisions between the two bins, $A$ and $B$, and how they induce low-frequency time dynamics in fluctuations of macroscopic flow parameters such as normalized overall stress, $\sigma_{xx}$, inside a shock.  
 We first develop a simple two energy-bin model similar to the Lotka-Volterra's predator-prey model~\citep{lotka1910,lotka1920,volterra1927fluctuations}.  
We then evaluate the collisions rate coefficients for different energy transfer processes using the DSMC particle distribution data, and show that instead of sustained particle oscillations, the solution to our ODE contains dampening terms that cause the periodic fluctuations to die out on time-scales an order of magnitude longer than the period of oscillation.

\subsection{The two-bin energy model}

\begin{figure}[H]
\begin{center}
\includegraphics[width=0.40\columnwidth]{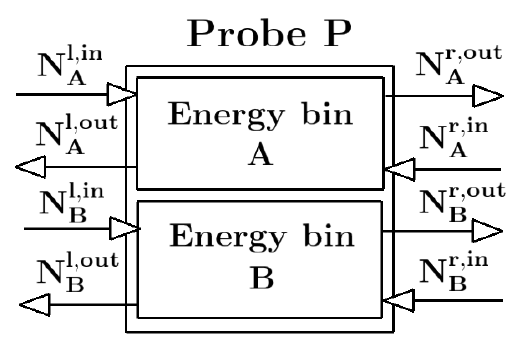}
\caption{A sketch denoting number of particles coming in and going out through the left and right boundary in a two-energy-bin model.}
\label{f:2BinModel}
\end{center}
\end{figure}

The dynamics of the number of particles in energy bins $A$ and $B$ may be written as,
\begin{equation} 
\centering
\begin{split}
\frac{dN_A}{dt} &= \left[\frac{dN_A}{dt}\right]_{coll} + F_A N_A\\
\frac{dN_B}{dt} &= \left[\frac{dN_B}{dt}\right]_{coll} + F_B N_B\\
\end{split}
\label{ODEModel}
\end{equation}
where the first and second terms on the right are the change in number of particles due to collisions and  fluxes, respectively.  
In addition, since the average number of  particles in energy bins $A$ and $B$ constitute 96.5\% of the total particles, we can assume them to be mutually exclusive and write,
\begin{equation} 
\centering
\begin{split}
\left[\frac{dN_A}{dt}\right]_{coll} &= -\left[\frac{dN_B}{dt}\right]_{coll}
\end{split}
\label{dNA_dt_equal_dNB_dt}
\end{equation}
The flux coefficient $F_j$ for  bin `$j$' is defined as the difference between the net average influx from the left boundary and outflux from the right boundary of the `$j$'-type particles per second divided by the average number of `$j$'-type particles,
\begin{equation} 
F_j  = \frac{(N_j^{l,in}-N_j^{l,out}) - (N_j^{r,out}-N_j^{r,in})}{\Delta t \langle N_j \rangle_t} \text{,\,\,\,\,\, j$\in$ \{A,B\}} 
\end{equation}
where, $\Delta t$ is the DSMC timestep, 
superscripts `$l$' and `$r$' refer to the left and right boundaries of the computational cell, respectively, and `$in$' and `$out$' refer to the incoming and outgoing particles, respectively, as shown in figure~\ref{f:2BinModel} for location $P$ and energy bins $A$ and $B$.
\vspace{\baselineskip}

Flux coefficients are required in the evaluation of equation~\ref{ODEModel}.  
Using the values of $A$ and $B$-type particles given in table~\ref{tab:NumberFlux}, flux values of $F_A$ and $F_B$ = -1,277,925.8 and  2,500,681.6~s$^{-1}$, respectively, at probe $P$ and zero in the freestream (as expected) are obtained.  
Note that at probe $P$, $F_A<0$ because on average more $A$-type particles  travel to a cell downstream  than  came in from the upstream, as also seen from figure~\ref{f:AvgParticles_ProbeP}.  The opposite  is true for type-$B$ particles, resulting in $F_B>0$.     
 This is synonymous with the fact that if there were no collisions, i.e., $\left[\frac{dN_A}{dt}\right]_{coll}=0$, the number of $A$ particles, those that mainly represent the subsonic part of the bimodal distribution, would decrease, while the number of 
$B$ particles, those that mainly represent the upstream hypersonic flow, would increase.

\begin{figure}
    \centering
    \sidesubfloat[]{\label{fa:zetaX_A_P}{\includegraphics[width=0.46\textwidth]{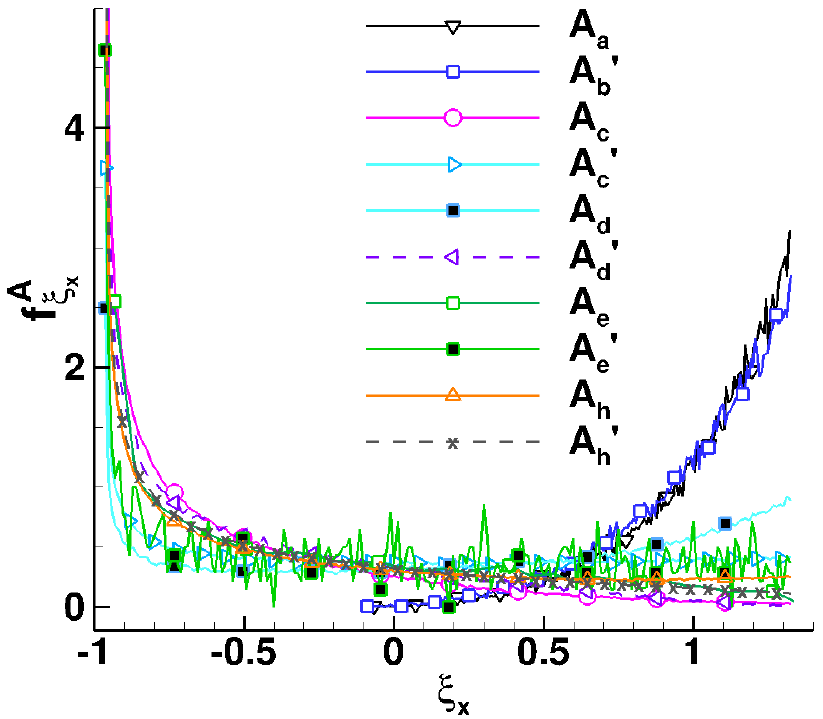}}}\hfill
    \sidesubfloat[]{\label{fb:zetaYZ_A_P}{\includegraphics[width=0.46\textwidth]{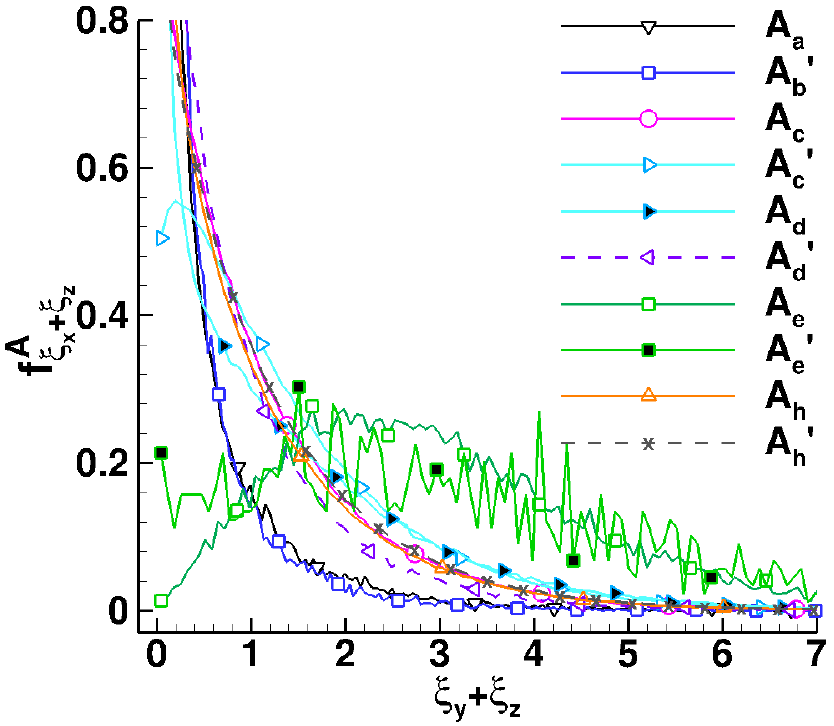}}}\,
    \sidesubfloat[]{\label{fc:zetaX_A_F}{\includegraphics[width=0.46\textwidth]{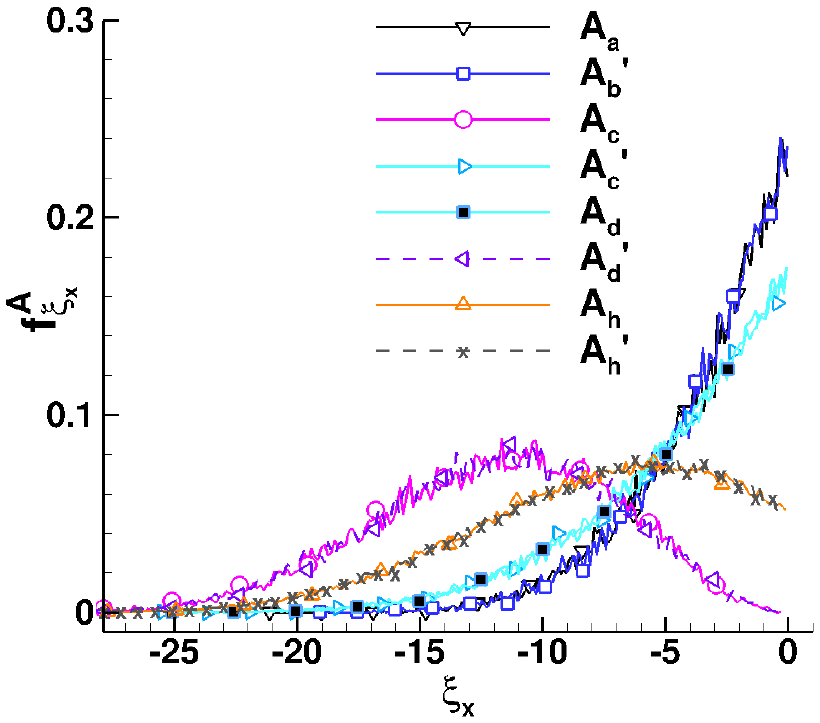}}}\hfill
    \sidesubfloat[]{\label{fd:zetaYZ_A_F}{\includegraphics[width=0.46\textwidth]{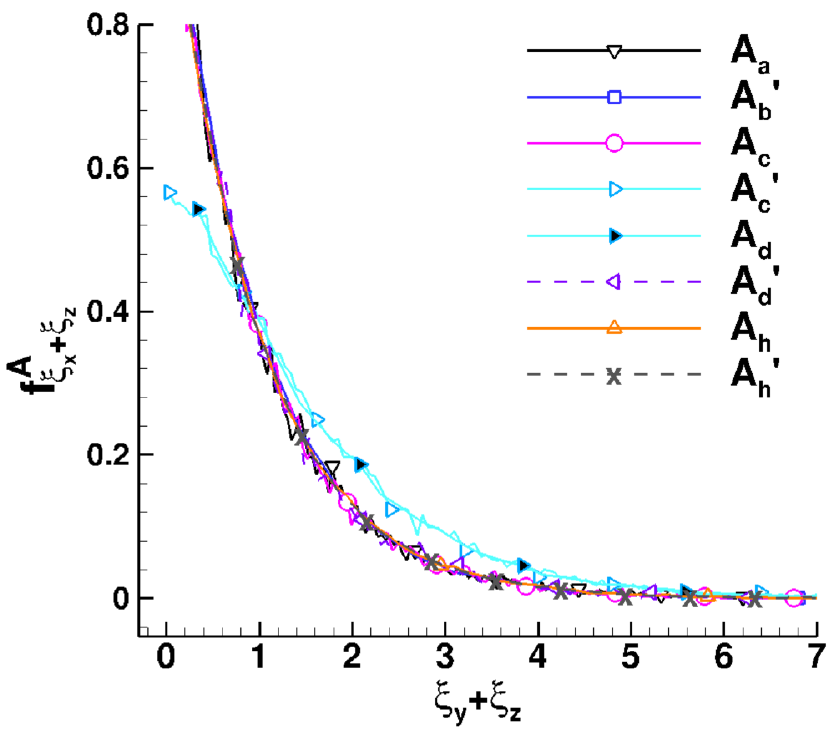}}}\,
    \caption{(\textit{a}) and (\textit{b})  show, at Probe $P$, the PDFs $f_{\xi_x}^A$ and $f_{\xi_y+\xi_z}^A$ of all sub-types of type-$A$ particles distinguished based on detailed collision processes listed in table~\ref{tab:collProc}. Note that $f_{\xi_x}^A$ is obtained by further refining the $X$-directional energy space of bin $A$ ($-0.86 < \xi_x < 1.33$) with a resolution of $\Delta\xi_x=0.0115$. Same resolution was used for the transverse energy space. (\textit{c}) and (\textit{d}) show the respective plots for probe $F$.}
    \label{f:binADist}
\end{figure}

\begin{table}[H]
\begin{threeparttable}
\caption{Fluxes of DSMC computational particles per timestep and computational cell area of $\num{1e-8}$~m$^2$.}
\label{tab:NumberFlux}       
\begin{tabular}{cccccccccc}
\hline\noalign{\smallskip}
\textbf{Probes} & $\mathbf{N_A^{l,in}}$ & $\mathbf{N_A^{l,out}}$ & $\mathbf{N_A^{r,out}}$ & $\mathbf{N_A^{r,in}}$ & $\mathbf{N_B^{l,in}}$ & $\mathbf{N_B^{l,out}}$ & $\mathbf{N_B^{r,out}}$ & $\mathbf{N_B^{r,in}}$    \\\hline
$P$ & 41.46 & 7.19 & 47.40 & 8.96 & 69.84 & 0.68 & 65.37 & 0.80 \\\
$F$ & 49.28  & 0 & 49.28 & 0.0 & 58.63 & 0.0 & 58.63 & 0.0 \\
\noalign{\smallskip}\hline\noalign{\smallskip}
\end{tabular}
\end{threeparttable}
\end{table}
To evaluate the collision terms in equation~\ref{ODEModel}, we need to identify collision processes that can cause a loss or gain of type-$A$ and -$B$ particles.
Such collision processes are listed in column two of table~\ref{tab:collProc}, where the {\it type} of $A$ particle is labeled based on the collision process that it undergoes, denoted by a subscript  `$i$', the relevant pre-and post-collisional states are specified where the latter is denoted by a primed superscript, and the
 rate coefficient for each {\it fundamental} collision process, $k_i$,  unless it is a ``compound'' rate which is denoted by a tilde.
The table shows that there are six fundamental processes (2nd column) and eight types of $A$ particles 
 that can cause a net change in the number of particles in energy bins $A$ and $B$.  
Using the DSMC-derived PDFs for all sub-types of type-$A$ particles of the normalized $X$-directional energy, $\xi_x=(v_x^2-u_x^2)\beta^2$, and total transverse energy, $\xi_y+\xi_z=(v_y^2+v_z^2)\beta^2$, denoted by $f_{\xi_x}^A$ and $f_{\xi_y+\xi_z}^A$, respectively (figure~\ref{f:binADist}), we can reduce the number of types of $A$ particles further as follows.
\vspace{\baselineskip}

Collision process $p_a$ describes a mechanism for type-$A_a$ particles in energy bin $A$ that collide with particles in energy bin $B$ and now belong to bin $B$ resulting in the loss of type-$A$ particles. 
Simultaneously, process $p_b$ describes the collision process of two bin $B$ particles that causes one of them to move to energy bin $A$ leading to a gain of type-$A$ particles, denoted by  $A_b'$.  
However,  figures~\ref{fa:zetaX_A_P} and~\ref{fb:zetaYZ_A_P} show that at location $P$ the energy distributions $f_{\xi_x}^A$ and $f_{\xi_y+\xi_z}^A$ for $A_a$ and $A_b$ particles are the same.
Therefore, we can group $A_b$ and $A_a$-type particles and write the second reaction $P_b$ as shown in column three of table~\ref{tab:collProc}.
The same holds true for probe $F$ as shown in figures~\ref{fc:zetaX_A_F} and~\ref{fd:zetaYZ_A_F}, however, the distributions different from those at probe $P$.

\begin{table}[H]
\normalsize
\begin{threeparttable}
\caption{Detailed and simplified collision processes.}
\label{tab:collProc}       
\begin{tabular}{ccccc}
\hline\noalign{\smallskip}
\textbf{Id} & \textbf{Detailed collision process}               &    \textbf{\makecell{Collision Process\\ after grouping\\ alike A-particles}}                   &    \textbf{\makecell{Simplified\\ collision Process} }                            \\
$\mathbf{i}$ & $\mathbf{p_i}$                                 &   $\mathbf{P_i}$                                          &     $\mathbf{Q_i}$                   \\\hline
a & \ch{A$_{a}$ + B   ->[ $k_{a}$ ] B         + B}              & \ch{A$_{a}$ + B   ->[ $k_{a}$ ] B         + B}              & \ch{A + B   ->[ $\tilde{k}_{a}$ ] B + B}      \\
b & \ch{B + B           ->[ $k_{b}$ ] A$_{b}'$ + B}              & \ch{B + B           ->[ $k_{b}$ ] A$_{a}$ + B}              & \ch{B + B   ->[ $k_{b}$ ] A + B}    \\
c & \ch{A$_{c}$ + B   ->[ $k_{c}$ ] A$_{c}'$ + A$_{c}'$}        & \ch{A$_{c}$ + B   ->[ $k_{c}$ ] A$_{c}'$ + A$_{c}'$}        & \ch{A + B   <->[ $\tilde{k}_{c}$ ] A$_{c}'$ + A$_{c}'$} \\
d & \ch{A$_{d}$ + A$_{d}$ ->[ $k_{d}$ ] A$_{d}'$ + B }          & \ch{A$_{c}'$ + A$_{c}'$ ->[ $k_{d}$ ] A$_{c}$ + B }         & -              \\
e & \ch{B + B ->[ $k_{e}$ ] A$_{e}'$ + A$_{e}'$}                & \ch{B + B ->[ $k_{e}$ ] A$_{e}'$ + A$_{e}'$}                & \ch{B + B ->[ $k_{e}$ ] A$_{e}'$ + A$_{e}'$}  \\
f & \ch{A$_{f}$ + A$_{f}$ ->[ $k_{f}$ ] B + B}                  & \ch{A$_{e}'$ + A$_{e}'$ ->[ $k_{f}$ ] B + B}                &   -                                                  \\\hline  
g$^1$ & \ch{A$_{g}$ + B   ->[ $k_{g}$ ] B + A$_{g}'$ }              & \ch{A$_{g}$ + B   ->[ $k_{g}$ ] B + A$_{g}$ }               &    -                                                  \\
h$^1$ & \ch{A$_{h}$ + A$_{h}$   ->[ $k_{h}$ ] A$_{h}'$ + A$_{h}'$}  & \ch{A$_{h}$ + A$_{h}$   ->[ $k_{h}$ ] A$_{h}$ + A$_{h}$}    &     -                      \\
i$^1$ & \ch{B + B   ->[ $k_{i}$ ] B + B}                            & \ch{B + B   ->[ $k_{i}$ ] B + B}                            &      -                                                \\
\noalign{\smallskip}\hline\noalign{\smallskip}
\end{tabular}
\begin{tablenotes}
\item $^1$ These processes cause no net change in the number of particles of energy bins $A$ and $B$.
\end{tablenotes}
\end{threeparttable}
\end{table}

We can also group all collisions between type-$A$ particles that do not lead to a net change in the energy distribution of particles within bin $A$ as process $p_h$ and see from figure~\ref{f:binADist} that the energy distributions of all such type-$A$ particles before and after collisions, $A_{h}$ and $A_{h}'$,  is the same at probe $P$ as well as $F$.  
Note that  at probe $F$, the distribution $f_{\xi_y+\xi_z}^A$ for $A_a$-type particles matches with the $A_h$-type, {\em i.e.,} in the freestream the transverse energy of particles taking part in processes $P_a$ and $P_b$ is the same as those in $P_h$.
The differences at probe $P$ indicate the role of transverse translational modes in the collision processes $P_a$ and $P_b$.
Similar to process $p_h$, process $p_g$,  is a bin-exchange process, which causes no net change in the number of particles in bins $A$ and $B$.
Particles simply swap energy bins by exchanging $X$-directional energies.  Using similar logic, additional simplifications can be made to the detailed collision processes, $p_i$,  to construct the  $P_i$ column of table~\ref{tab:collProc} based on the energy distribution functions at locations $P$ and $F$.  The details of these analyses may be found in appendix~\ref{app:simplification}.  
\vspace{\baselineskip}

To complete the dynamical model for particles moving between energy bins $A$ and $B$, an expression is required for the net change in the number of particles in energy bin $A$ due to collisions from all possible processes related to $A$-type particles, {\em i.e.,}
\begin{equation} 
\centering
\begin{split}
\left[\frac{dN_A}{dt}\right]_{coll} &= \left[\frac{dN_{A_{a}}}{dt}\right]_{coll} + 
                                       \left[\frac{dN_{A_{c}}}{dt}\right]_{coll} +
                                       \left[\frac{dN_{A_{c}'}}{dt}\right]_{coll} + 
                                       \left[\frac{dN_{A_{e}'}}{dt}\right]_{coll} 
\end{split}
\label{dNA_sum}
\end{equation}
where $N_{A_{a}}$, $N_{A_{c}}$, $N_{A_{c}'}$, and $N_{A_{e}'}$ are the number of particles of $A_a$, $A_c$, $A_{c}'$, and $A_{e}'$ types, respectively.
Using standard rate equation formulations for elementary reactions \citep[see][sec. 16.15]{andersonModern}, the terms on the right hand side can be expressed as,
\begin{equation} 
\centering
\begin{split}
\left[\frac{dN_{A_{a}}}{dt}\right]_{coll} &= -k_a N_{A_{a}} N_{B} + k_b N_{B}^2 \\
\left[\frac{dN_{A_{c}}}{dt}\right]_{coll} &= -k_c N_{A_{c}} N_{B} + k_d N_{A_{c}'}^{2} \\
\left[\frac{dN_{A_{c}'}}{dt}\right]_{coll} &= 2 k_c N_{A_{c}} N_{B} - 2 k_d N_{A_{c}'}^2 \\
\left[\frac{dN_{A_{e}'}}{dt}\right]_{coll} &= 2 k_e N_{B}^2 - 2 k_f N_{A_{e}'}^2 \\
\end{split}
\label{dNA_subspecies}
\end{equation}
By substituting equation~\ref{dNA_subspecies} in~\ref{dNA_sum}, we obtain,
\begin{equation} 
\centering
\begin{split}
\left[\frac{dN_A}{dt}\right]_{coll} &= -k_a N_{A_{a}} N_{B} + k_b N_{B}^2 + k_c N_{A_{c}} N_{B} - k_d N_{A_{c}'}^{2} \\
                                    &+ 2 k_e N_{B}^2 - 2 k_f N_{A_{e}'}^2\\
\end{split} 
\label{dNA_sum_substituted}
\end{equation}
As shown in the  appendix~\ref{app:simplification}, the conditions in the 1-D shock allow us to make additional simplifications to equation~\ref{dNA_sum_substituted},  so that using equations~\ref{ODEModel},~\ref{dNA_dt_equal_dNB_dt}, and~\ref{dNA_sum_simplified} we obtain the final system of ordinary differential equations used to study perturbation dynamics,
\begin{equation} 
\centering
\begin{split}
\frac{dN_A}{dt} &= (\tilde{k}_c-\tilde{k}_a) N_{A} N_{B} + (k_b+2k_e) N_{B}^2 + F_A N_A \\
\frac{dN_B}{dt} &= -(\tilde{k}_c-\tilde{k}_c) N_{A} N_{B} - (k_b+2k_e) N_{B}^2 + F_B N_B
\end{split} 
\label{ODEModel_Final}
\end{equation}

\subsection{Evaluation of rate coefficients}
Finally, to solve the dynamical system of equations we need to evaluate the rate coefficients in equation~\ref{ODEModel_Final}. 
A summary of the expressions for the rate coefficients used in equation~\ref{ODEModel_Final} and their values are presented in 
 table~\ref{tab:RateCoeff} for probes $P$ and $F$  based on the DSMC collision data collected during the simulation, summarised in table~\ref{tab:DSMCNCols}.  
Note that $\langle C_{i} \rangle_t$ is the average number of DSMC collisions per timestep per cell volume that take part in the $i^{th}$ detailed collision process of particle types $R_1$ and $R_2$ (see. equation~\ref{RateCoeff_ki}).  
 The rate coefficients $k$/($s^{-1}$) shown in table~\ref{tab:RateCoeff} can be converted to the traditional units of $K/(m^3.s^{-1})$ by multiplying them with the volume of computational cell, $\Delta v=\num{1e-12}~m^3$, and dividing by the parameter, FNUM=$\num{1e7}$.
By this conversion, it can be easily shown that the rates $\tilde{K}_{a}$, $K_b$ range from $1.8-\num{5.0e-17}$~m$^3$.s$^{-1}$ at probes $P$ and $F$.   

\begin{table}[H]
\centering
\normalsize
\begin{threeparttable}
\caption{Rate coefficients per second for collision processes defined in table~\ref{tab:collProc}.}
\label{tab:RateCoeff}       
\begin{tabular}{cccc}
\hline\noalign{\smallskip}
\textbf{Rate coefficient}      & \textbf{Formula}                                         & \textbf{Probe P} & \textbf{Probe F}\\\hline
$\tilde{k}_a$        &${\langle C_{a}\rangle_t}\big/{\Delta t \langle N_{A} \rangle_t \langle N_B\rangle_t}$                        & 188.9  & 497.0 \\
\addlinespace[3pt]
$k_b$                  &${\langle C_{b}\rangle_t}\big/{\Delta t \langle N_B \rangle_t^2} $                                            & 417.5  & 513.5\\
\addlinespace[3pt]
$\tilde{k}_c$        &${\left(\langle C_c  \rangle_t -\langle C_d \rangle_t\right)}\big/{\Delta t \langle N_{A} \rangle_t \langle N_B\rangle_t}$ & 1843.5 & 0\\
\addlinespace[3pt]
$k_e$                  &${\langle C_e  \rangle_t}\big/{\Delta t \langle N_B\rangle_t^2} $                                             & 75.0   & 0 \\
\noalign{\smallskip}\hline\noalign{\smallskip}
\end{tabular}
\end{threeparttable}
\end{table}

\begin{table}[H]
\begin{threeparttable}
\caption{Average number of DSMC collisions per timestep per computational cell volume, $\langle C_{i} \rangle_t$.}
\label{tab:DSMCNCols}       
\begin{tabular}{cccccccccc}
\hline\noalign{\smallskip}
Probes & $\mathbf{\langle C_a \rangle_t }$ & $\mathbf{\langle C_b \rangle_t}$ & $\mathbf{\langle C_{c} \rangle_t}$ & $\mathbf{\langle C_{d} \rangle_t}$ & $\mathbf{\langle C_{e} \rangle_t }$ & $\mathbf{\langle C_{f} \rangle_t }$ & $\mathbf{\langle C_{g} \rangle_t}$  & $\mathbf{\langle C_{h} \rangle_t}$ & $\mathbf{\langle C_{i} \rangle_t}$       \\\hline
 $P$   & 0.377 & 0.468 & 5.074 & 1.396 & 0.084 & 0.0079 & 2.980 & 8.849 & 1.481\\
 $F$   & 0.375 & 0.374 & 0.382 & 0.383 & 0.0 & 0.0 & 1.028 & 0.893 & 0.887\\
\noalign{\smallskip}\hline\noalign{\smallskip}
\end{tabular}
\end{threeparttable}
\end{table}

Note that in the freestream, $\langle C_{a} \rangle_t \approx \langle C_{b}\rangle_t$, due to detailed balance in a local equilibrium condition, which dictates that each detailed collision must be balanced by its inverse collision process~\citep[see][pg.38-39 and 42]{vincenti1965introduction}.
Furthermore, the net rate, $\tilde{k}_{c}$ is nearly zero in the freestream because again by principle of detailed balance, $\langle C_{c} \rangle_t \approx \langle C_{d} \rangle_t$, as shown in table~\ref{tab:DSMCNCols}.
On the other hand, in the shock, a factor of 3.63 difference between the average values of these two types of collisions results in $\tilde{K}_{c}=\num{1.83e-16}$~m$^3$.s$^{-1}$.
Also, from table~\ref{tab:DSMCNCols} it can be seen that both $\langle C_{e} \rangle_t$ and $\langle C_{f} \rangle_t$ are zero in the freestream, which means that these processes take place only in conditions of translational nonequilibrium.
As a result, the rate $K_e$ is zero in the freestream, whereas in the shock it is two-orders of magnitude slower, $\num{7.5e-18}$~m$^3$.s$^{-1}$,   than $\tilde{K}_{c}$. 
\vspace{\baselineskip}

The rate coefficients given in table~\ref{tab:RateCoeff} can be compared with theoretical estimate of a rate coefficient,  $K_{th}$,   for a forward collision process of  a quasi-equilibrium gas \citep[see][pg. 213-216]{vincenti1965introduction}, which is defined as,
$$\ch{R$_1$ + R$_2$ ->[ $K_{th}$ ] S$_1$ + S$_2$}.$$
The rate of change of $R_1$-type particles per unit volume due to collisions is written as,
$$\left[\frac{dn_{R_1}}{dt}\right]_{coll} = K_{th} n_{R_1} n_{R_2}$$
where $n_{R_1}$ and $n_{R_2}$ are number densities of $R_1$ and $R_2$ type particles, respectively.
For a gas following the VHS molecular model, the rate coefficient $K_{th}$ can be expressed as,
\begin{equation}
\centering
\hspace{-21.2em}
K_{th} = \frac{1}{\varepsilon} \langle \sigma \rangle_t \langle g_r \rangle_t \gamma 
\label{RateCoeffTheory2}
\end{equation}
where
\begin{equation}
\centering
\begin{split}
\langle \sigma \rangle_t  &=  \pi d_{r}^2 \left(\frac{2\kappa_b T_{r}}{m_r \langle g_r \rangle_t}\right) \{\Gamma(2.5-\omega)\}^{-1}\\ 
\langle g_r \rangle_t                &=  \left(\frac{8 \kappa_b \langle T_{tr} \rangle_t }{\pi m_r}\right)^{\frac{1}{2}}\\
\gamma &= \left[\frac{1}{\Gamma\left(\frac{5}{2}-\omega\right)} \left(\Gamma\left(\frac{5}{2}-\omega, \frac{E_{a}}{\kappa_b \langle T_{tr} \rangle_t}\right) - \frac{E_{a}}{\kappa_b \langle T_{tr} \rangle_t}\Gamma\left(\frac{3}{2}-\omega, \frac{E_{a}}{\kappa_b \langle T_{tr} \rangle_t} \right) \right)\right]\\
E_{a} &= \frac{1}{2} m_r \langle g_r \rangle_t^2 
\end{split}
\nonumber
\end{equation}
where $\langle \sigma \rangle_t$ is the average equilibrium collision cross-section for the VHS gas, $\langle g_r \rangle_t$ is the time-averaged relative velocity, $\gamma$ is the fraction of collisions for which the relative translational energy along the line of centers of colliding molecules exceeds the activation energy of $E_a$~\citep[see][sec 6.2 and equation 4.72]{bird:94mgd}, and $\Gamma$ is the gamma function. 
Also, $m_r$ is the reduced mass~\citep[see][equation 2.7]{bird:94mgd}, $d$ is the VHS molecular diameter~\citep[see][equation 4.63]{bird:94mgd}, and $\varepsilon$ is a symmetry factor equal to two because R$_1$=R$_2$.
Table~\ref{tab:RateCoeffTheory} shows that the estimated parameters in equation~\ref{RateCoeffTheory2} and average rate coefficients at probes $P$ and $F$ located inside the shock and the freestream, respectively, are consistent with the values presented in table~\ref{tab:RateCoeff}.

\begin{table}[H]
\centering
\normalsize
\begin{threeparttable}
\caption{Parameters used in equation~\ref{RateCoeffTheory2} for the theoretical estimate of average rate coefficient.}
\label{tab:RateCoeffTheory}       
\begin{tabular}{cccc}
\hline\noalign{\smallskip}
\textbf{Parameters}      & \textbf{At Probe P}            & \textbf{At Probe F} \\\hline
$\langle T_{tr} \rangle_t$/(K)        & 10,440            & 710              \\
$\langle g_r \rangle_t$/(m.s$^{-1}$)  & 3326.2            & 867.6           \\
$\langle \sigma \rangle_t$/(m$^2$)  & \num{1.81e-19} & \num{4.16e-19} \\
$E_{a}$/(J)     & \num{1.83e-19}  & \num{1.25e-20}           \\
$\gamma$          & 0.351            & 0.351\\
$K_{th}$$^1$/(m$^3$.s$^{-1}$)         & \num{1.05e-16}  & \num{6.33e-17} \\
\noalign{\smallskip}\hline\noalign{\smallskip}
\end{tabular}
\end{threeparttable}
\end{table}

\section{Dynamics of Energy Fluctuations Inside a Shock using the Two-Energy-Bin Model}~\label{sec:Dynamics}
\begin{figure}[H]
    \centering
    \sidesubfloat[]{\label{f:ODESolnP}{\includegraphics[width=0.46\textwidth]{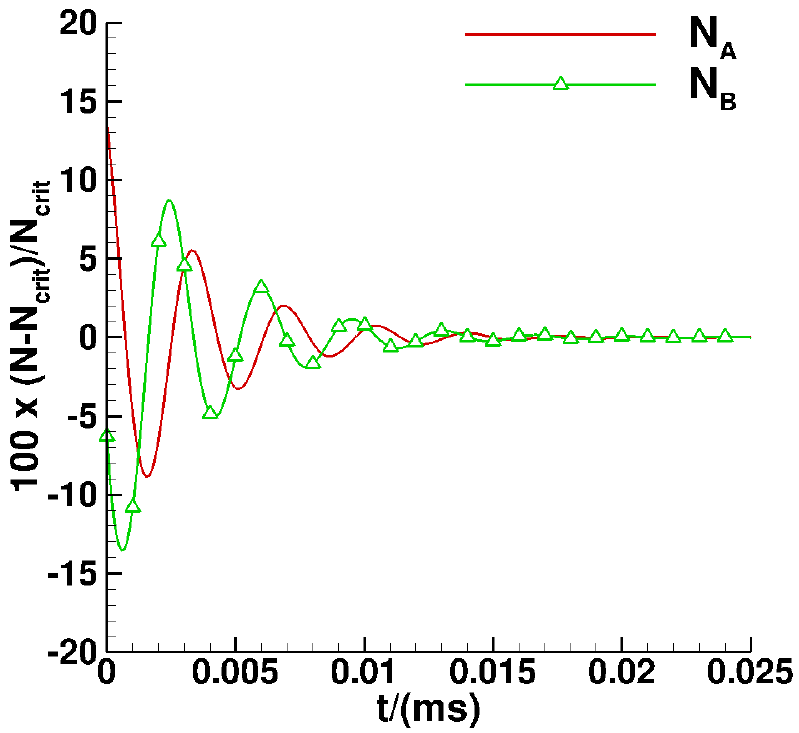}}}\hfill
    \sidesubfloat[]{\label{f:ODESolnF}{\includegraphics[width=0.46\textwidth]{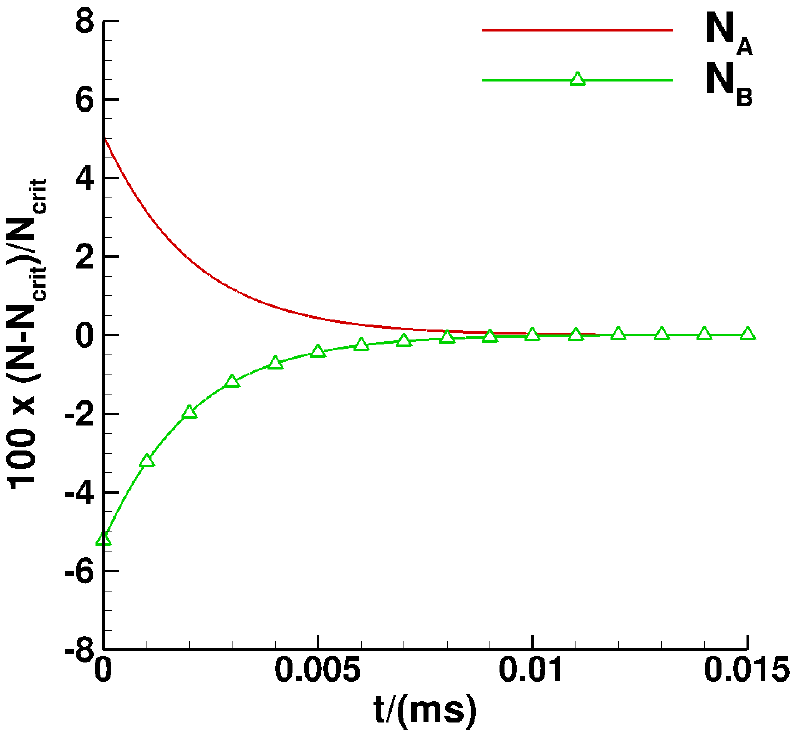}}}\,
    \caption{(\textit{a}) and (\textit{b}) show the solution of the two-energy-bin ODE model at probes $P$ and $F$, respectively. The initial values of $N_A$ and $N_B$ are taken to be +13.5 and -6.3\% of the critical values, respectively, at probe $P$, and +5.4 and -4.8\% of the critical values, respectively, at probe $F$. These values are three times the percent standard deviation observed from the DSMC simulation in the ratios of $N_A/\langle N_A \rangle_t$ and $N_B/\langle N_B \rangle_t$.}
    \label{f:ODESoln}
\end{figure}
\begin{figure}[H]
    \centering
    \sidesubfloat[]{\label{f:StreamplotP}{\includegraphics[width=0.46\textwidth]{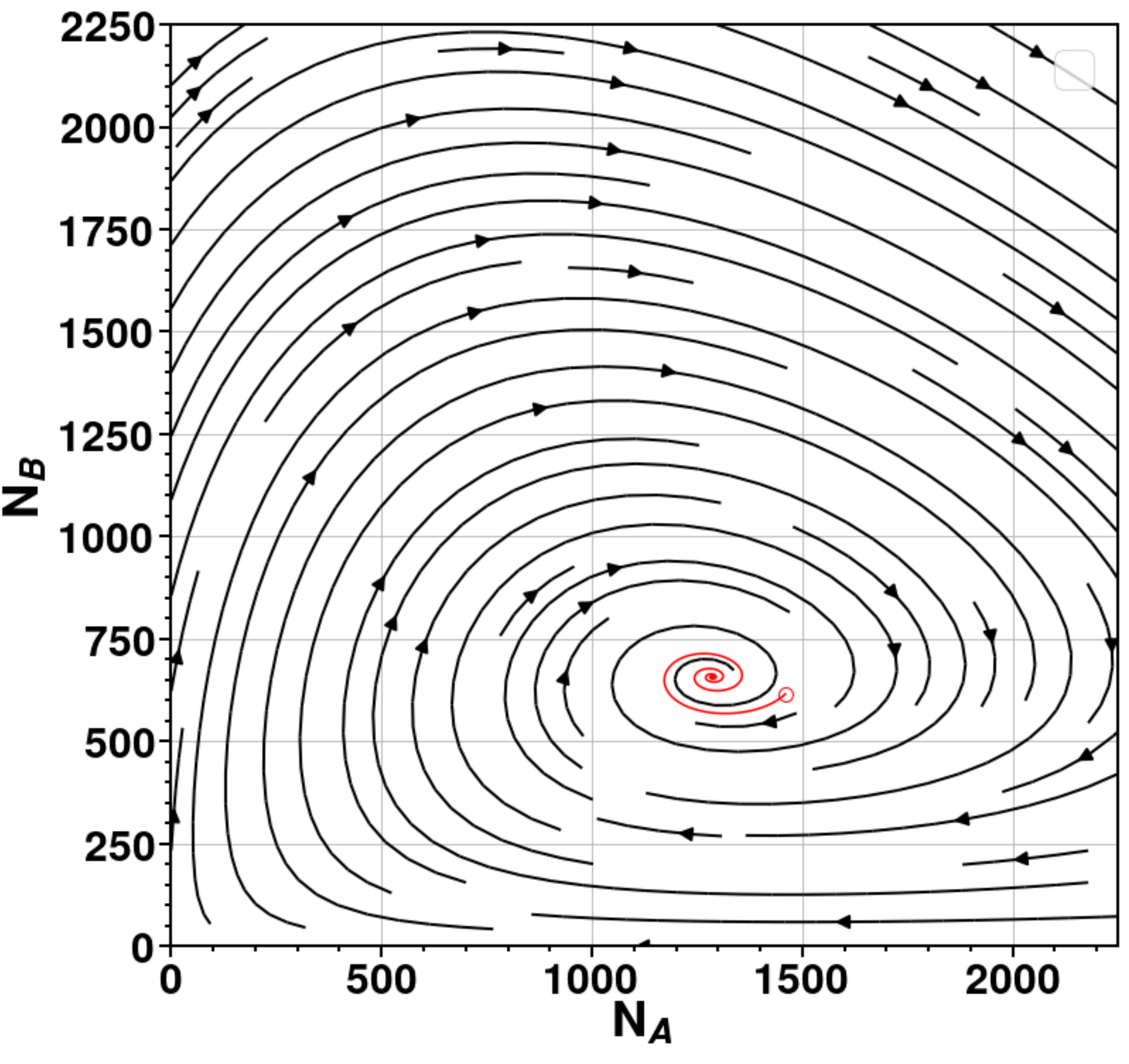}}}\hfill
    \sidesubfloat[]{\label{f:StreamplotF}{\includegraphics[width=0.46\textwidth]{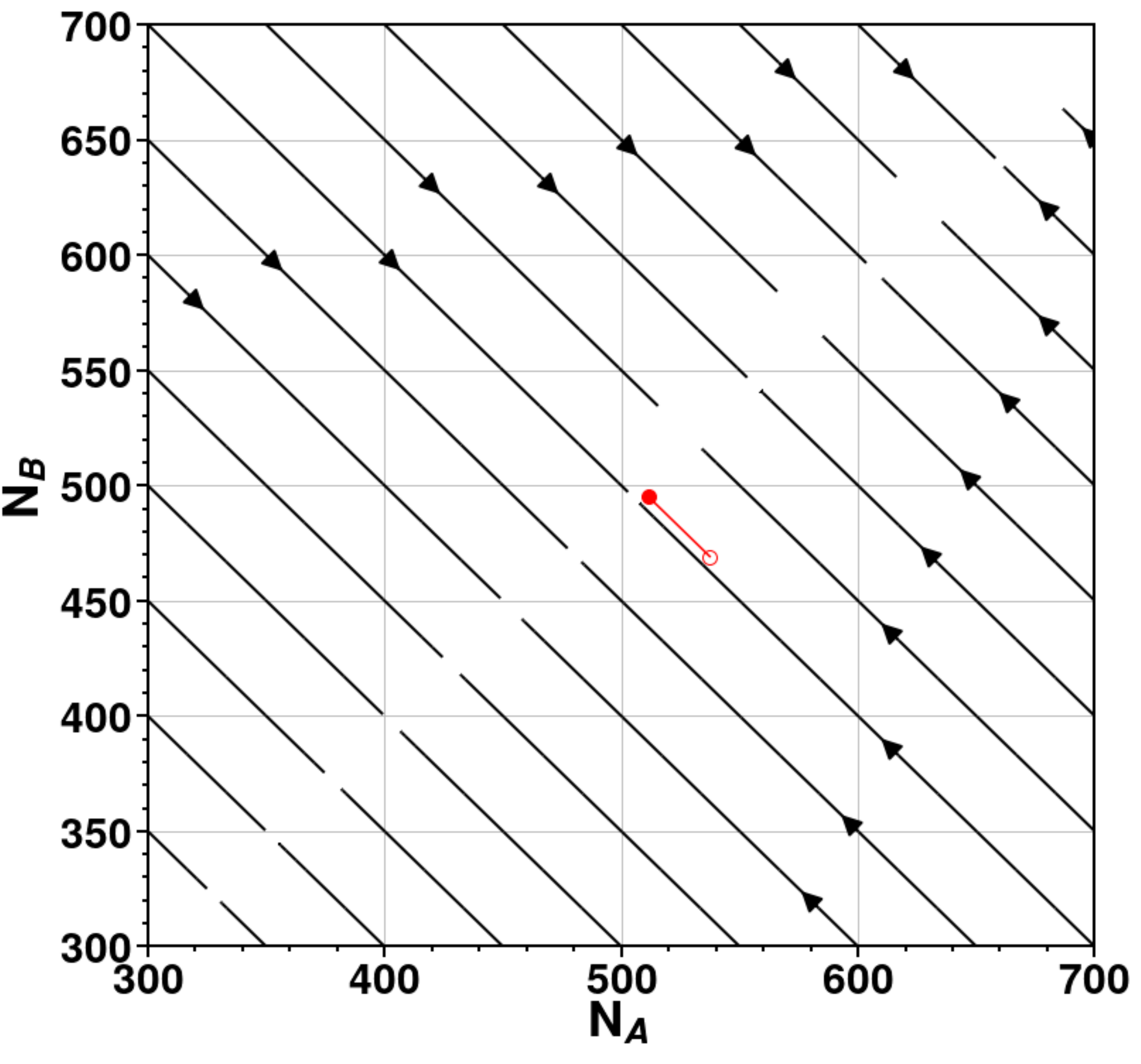}}}\,
    \caption{(\textit{a}) and (\textit{b}) show the streamlines of the normalized growth rate vector defined in equation~\ref{GrowthRateField} along with an overlaid path-line of the instantaneous solution ($N_A,N_B$) shown in figure~\ref{f:ODESoln} at probes $P$ and $F$, respectively. The starting and end points of the path-lines are denoted by hollow and filled circles.}
    \label{f:Streamplot}
\end{figure}

The numerical solution of the ODE system in equation~\ref{ODEModel_Final} at probes $P$ and $F$ is shown in figure~\ref{f:ODESoln} as a function of time and in figure~\ref{f:Streamplot} as path-lines in the phase space of $N_A$ versus $N_B$.
 Starting with location $P$, 
figure~\ref{f:ODESolnP} shows that the solution converges to the non-zero critical values of $N_{A,crit}$ and $N_{B,crit}$, obtained by setting $dN_A/dt=dN_B/dt=0$ which are listed in table~\ref{tab:ODESoln}.  
There is 18.2 and 7.5\% difference between the critical values and the average number of DSMC particles per computational cell volume in bins $A$ and $B$ at probe $P$, which may be attributed to the simplifications of the two-energy-bin model discussed above.  Yet, as will be shown, this simple model reveals an order of magnitude disparity in frequencies at probes $P$ and $F$. 
Figure~\ref{f:StreamplotP} shows the streamlines of a vector field defined by the normalized growth rate vector $\boldsymbol{G}=[G_{N_A}, G_{N_B}]^T$ with components,
\begin{equation} 
G_{N_i} = \frac{1}{\norm{\boldsymbol{G}}} \frac{dN_i}{dt}, \ \ i = A \ \ {\rm or} \ \ B \ \ {\rm and} \ \ \norm{\boldsymbol{G}} = \sqrt{\left(\frac{dN_A}{dt}\right)^2 + \left(\frac{dN_B}{dt}\right)^2}\
\label{GrowthRateField}
\end{equation}
We can see from the formation of a stable spiral that the critical point does not depend on the initial values.  The figure shows that the streamlines are tangent to the direction of the maximum growth vector at a given location of ($N_A, N_B$) and show eventual decay to the stable critical point of 1245 and 681.  
The overlaid path-line in figure~\ref{f:StreamplotP} describes these dynamics for the instantaneous solution shown in figure~\ref{f:ODESolnP}.

The Jacobian matrix of equation~\ref{ODEModel_Final} is given by,
\begin{equation} 
\centering
\begin{split}
J(N_A,N_B) &=
    \begin{bmatrix}
    (\tilde{k}_c-\tilde {k}_a) N_B + F_A & (\tilde {k}_c-\tilde{k}_a) N_A + 2 (k_b+2k_e) N_B \\ 
    -(\tilde{k}_c-\tilde {k}_a) N_B      & -(\tilde {k}_c-\tilde{k}_a) N_A - 2 (k_b+2k_e) N_B + F_B \\ 
    \end{bmatrix}  \\
\end{split} 
\label{Jacobian}
\end{equation}
The eigenvalues of matrix $J(N_{A,crit}, N_{B,crit})$ are listed in table~\ref{tab:ODESoln}.
At probe $P$ they are complex conjugates with a negative real part in the shock that gives the rate of decay from the initial to critical values.  
This is consistent with figure~\ref{f:ODESolnP}, which shows that at 0.01, 0.015, and 0.02~ms, the solution converges to 1, 0.3, and 0.07\% of the critical values, respectively.
These percentage values correspond to an approximate difference between the critical and instantaneous values of the number of $A$-type particles of 13, four, and one, respectively.
\vspace{\baselineskip}

More importantly, these results suggest that inside the shock, the number of particles, which are subjected to recurrent microscopic thermal fluctuations, always decay with time-scales on the order of 0.01-0.02~ms with corresponding frequencies on the order of 100 to 50~kHz, close to the weighted-average of the low-frequency broadband of 37.5~kHz observed in figure~\ref{f:PSD_ProbeP} from the PSD analysis.
Furthermore, the DSMC residual of the instantaneous data of overall stress, shown in figure~\ref{f:ProbeDataStresses}, and the instantaneous particles in the energy bins $A$ and $B$, shown in figure~\ref{f:NaNb_ProbeP}, are a result of superposition of responses to recurrent low-frequency microscopic fluctuations.
Additionally, the imaginary part of the pair of eigenvalues listed in table~\ref{tab:ODESoln} at probe $P$ indicates an oscillating behavior with a time period of $2\pi/\operatorname{Im}(L_1)$,  corresponding to a frequency of 281~kHz. This value is amazingly close to the 
 286~kHz frequency seen in figure~\ref{f:ProbeDataStresses} of the instantaneous residual of overall stress and frequencies close to 300~kHz outside of the low-frequency broadband, seen in figure~\ref{f:PSD_ProbeP} from the PSD analysis. 
Therefore, even with the underlying simplifications of the two-energy-bin model, it is able to predict the existence of the low-frequencies that originate due to the interaction in the  number of particles between the two modes of the bimodal PDF of particle energies in the presence of translational nonequilibrium in a 1-D shock.

\begin{table}[H]
\centering
\normalsize
\begin{threeparttable}
\caption{Dynamics of the ODE at probes $P$ and $F$.}
\label{tab:ODESoln}       
\begin{tabular}{cccc}
\hline\noalign{\smallskip}
\textbf{Solution} & \textbf{Probe P} & \textbf{Probe F}\\\hline
$N_{A,crit}$            & 1286    & 511.5 \\
$N_{B,crit}$            & 657.2   & 495 \\
$L_1$               & -281 762 + 1 765 303.3$j$   & 0.0\\
$L_2$               & -281 762 - 1 765 303.3$j$   & -499 976\\
\noalign{\smallskip}\hline\noalign{\smallskip}
\end{tabular}
\end{threeparttable}
\end{table}

Turning to probe $F$ in the freestream, we want to demonstrate that the two-bin energy model is able to predict different dynamics than observed in the shock.  
Since the flux coefficients $F_A$ and $F_B$ and the rate coefficients $\tilde{k}_c$ and $k_e$ are zero, we obtain,
\begin{equation} 
\centering
\begin{split}
\frac{N_{A,crit}}{N_{B,crit}}= \frac{k_b}{\tilde{k}_a}\\
\end{split} 
\label{CriticalPoints_F}
\end{equation}
which is satisfied by an infinite number of critical points corresponding to different freestream conditions.  
That is, the number densities are different but the temperature is the same since all solutions have the same shape of the PDF shown in figure~\ref{f:fex_P} and therefore, the same cross-correlation coefficient shown in figure~\ref{f:CCF_ProbeF}.
In this case, the solution depends on the initial values of $N_A$ and $N_B$, and converges to a critical point that satisfies equation~\ref{CriticalPoints_F}.
Figures~\ref{f:ODESolnF} and~\ref{f:StreamplotF} show the solution of equation~\ref{ODEModel_Final} as a function of time for a set of  initial values of ($N_A, N_B$) and a streamplot along with an overlaid path-line of the solution, respectively. 
In figure~\ref{f:StreamplotF}, the dependence on the initial values can be clearly seen. 
For the given conditions, the solution converges to critical values of $N_{A,crit}=511.5$ and $N_{B,crit}=495$, which are within 0.5\% of the average number of DSMC particles per computational cell volume, of $\langle N_A \rangle_t=509.9$ and $\langle N_B \rangle_t=492.8$, and satisfy the ratio of 1.033 given by equation~\ref{CriticalPoints_F}.
\vspace{\baselineskip}

For the freestream conditions, there are two real eigenvalues, as shown in table~\ref{tab:ODESoln} with the zero eigenvalue indicating again a 
 non-unique solution of equation~\ref{CriticalPoints_F}.
The negative eigenvalue of -499,976 gives the linear rate of decay, the inverse of which gives the characteristic time-scale of  2~$\mu$s.
This is consistent with the PSD results discussed in section~\ref{sec:Perturbations}, which revealed that 40\% of the total spectral energy contained within a band of 93 to 443~kHz including the peak, corresponding to an order of magnitude higher time-scales than the mean-collision-time of 0.284~$\mu$s.
More importantly, in the freestream the model predicts the absence of an order of magnitude lower dominant frequencies having significantly large energy, which appear to be unique to the region of strong nonequilibrium inside a shock.
\vspace{-1em}
\section{Strouhal Numbers at Various Input Conditions}~\label{sec:Scaling}
\begin{figure}[H]
    \sidesubfloat[]{\label{f:FreqVsMach}{\includegraphics[width=0.49\textwidth]{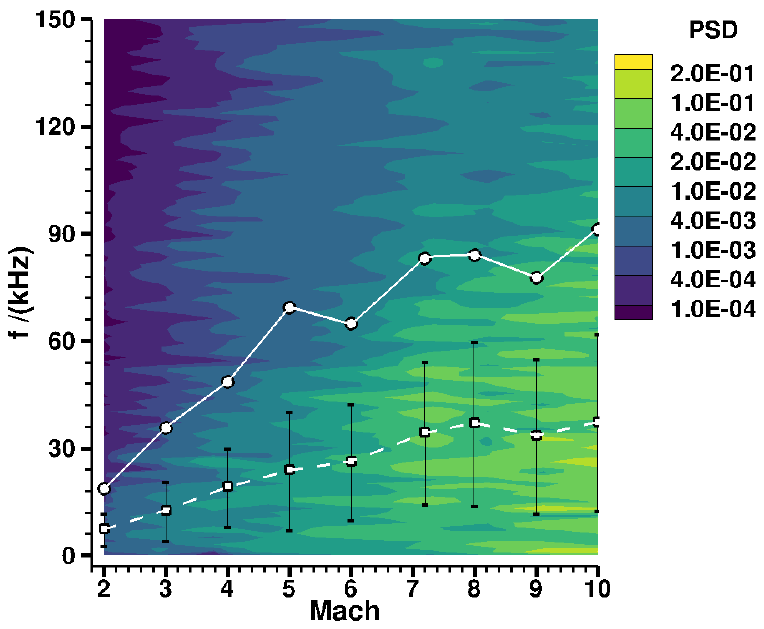}}}
    \sidesubfloat[]{\label{f:NCEs}{\includegraphics[width=0.43\textwidth]{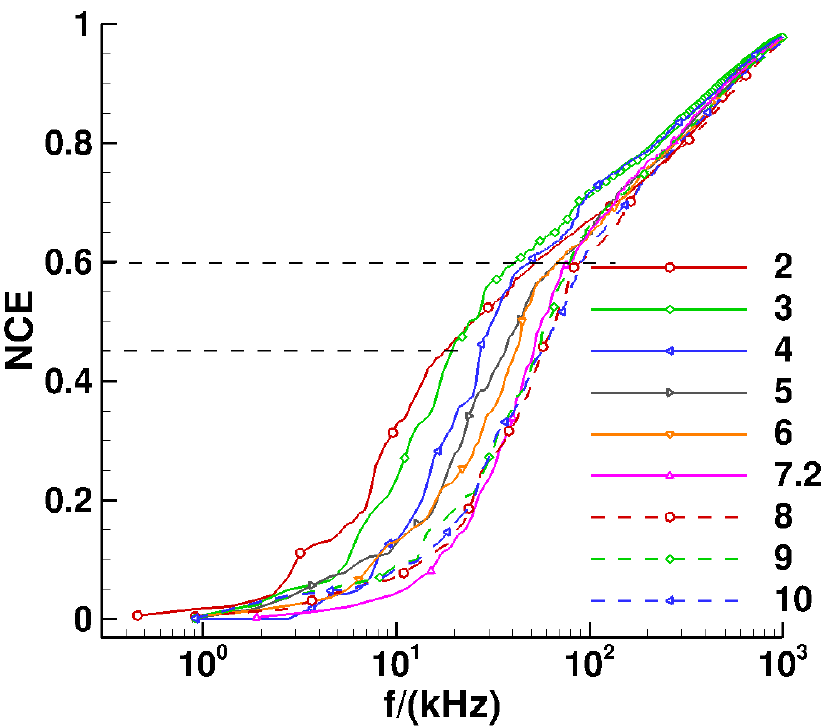}}}
    \caption{(\textit{a}) The contours of PSD obtained by interpolating the pressure spectrum at the center of the shock at each Mach number.  For interpolation, the inverse-distance algorithm in the \citet{Tecplot} software is used with default parameters (exponent=3.5, point selection=Octant, Number of points=8). The overlaid while solid line shows the demarcation boundaries at 60\% of the total spectral energy for Mach numbers from 3 - 10 and 45\% for Mach 2, and the dashed line shows the weighted-average of the frequencies in this spectral region with $\pm$1 standard deviation. (\textit{b}) NCE of the PSD for different Mach numbers. }
    \label{f:PSD_AllMach}
\end{figure}

With the observation of dominant low-frequency perturbations of macroscopic flow parameters inside the shock, a natural question arises as to whether one can define 
 a nondimensional Strouhal number,
 \begin{equation} 
St = \frac{f L_s}{u_{x,1}}
\label{StrouhalNumber}
\end{equation}
that would be constant for different shock strengths.  
We propose to define the time-scale as that required for the flow to traverse a distance equal to the shock-thickness with the upstream bulk velocity.
To justify this time-scale, cases ranging from Mach two to 10 were run by varying the upstream bulk velocity but keeping the upstream temperature constant ($T_{tr,1}$=710~K).
Starting with the selection of frequency, $f$, 
the PSD of the instantaneous pressure spectrum at the location of the maximum gradient in the shock for different Mach numbers is shown in figure~\ref{f:FreqVsMach}.
A broadband of low-frequencies is seen, the boundary of which is defined by the inflection point in the NCE, shown in figure~\ref{f:NCEs}.
\vspace{\baselineskip}

\begin{figure}[H]
    \centering
    \sidesubfloat[]{\label{f:ShockT}{\includegraphics[width=0.45\textwidth]{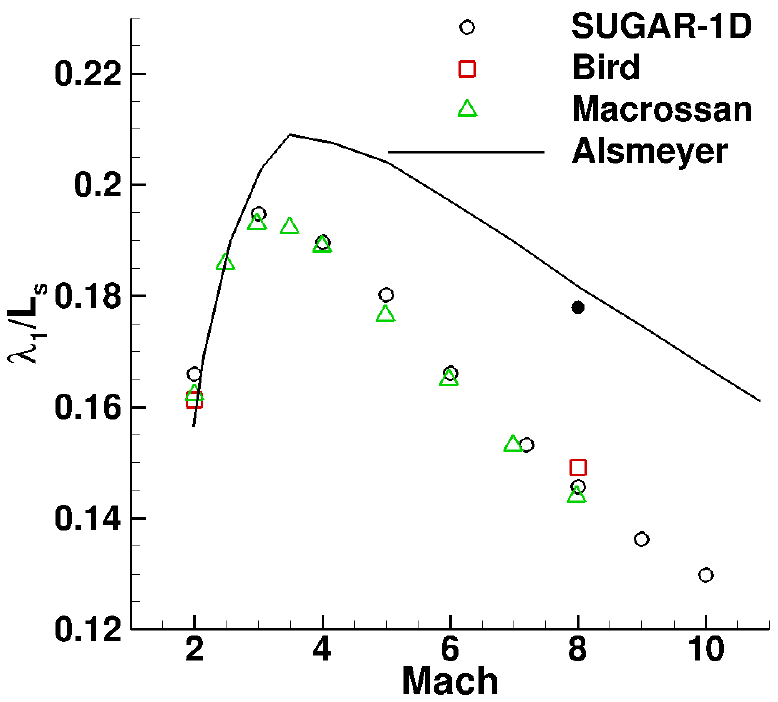}}}\hfill
    \sidesubfloat[]{\label{f:St}{\includegraphics[width=0.49\textwidth]{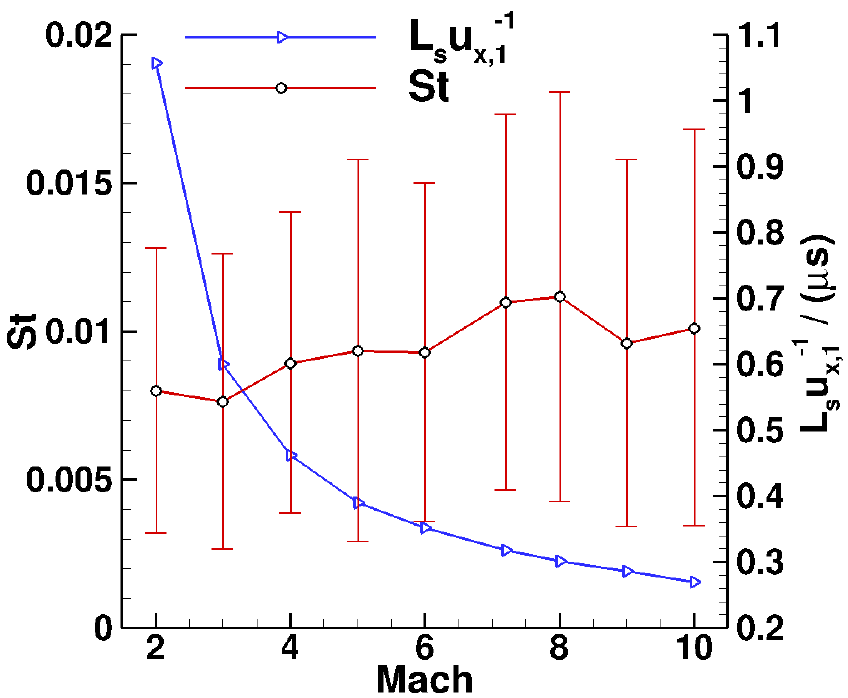}}}\,
    \caption{(\textit{a}) Reciprocal of the density-gradient shock-thickness normalized by the upstream mean-free-path as a function of Mach number. The SUGAR 1-D DSMC results use a viscosity index of $\omega=0.81$ (open symbols) and $\omega=0.75$ (filled circular symbol) at Mach 8.
 (\textit{b})  $L_s u_{x,1}^{-1}$ and Strouhal number as a function of Mach number. The standard deviation in Strouhal number is based on the standard deviation in weighted-average frequency shown in figure~\ref{f:FreqVsMach}.  }
    \label{f:StrouhalDef}
\end{figure}

The characteristic length, $L_s$, defined by the density-gradient shock-thickness is calculated as the overall density change divided by the maximum density gradient~\citep[for example, see][chapter X, sec. 9]{vincenti1965introduction} and is shown in figure~\ref{f:ShockT}.
Note that $\lambda_1$, the upstream mean-free-path, is obtained through the VHS model with a viscosity index of $\omega=0.81$.
The SUGAR-1D DSMC shock-thickness values match within 2\% with the DSMC calculations of \citet{macrossan2003viscosity} and \citet[see][chapter 12, pg. 294]{bird:94mgd} for the same viscosity index. 
Note that their results are scaled by 76\% to account for the differences in $\lambda_1$, which they defined based on the hard-sphere model~\citep{bird:94mgd}.
The noticeable discrepancy between the SUGAR-1D DSMC results for $\omega=0.81$ with the experiments of \citet{alsmeyer1976density} is due to the choice of viscosity index, while qualitatively the variation of shock-thickness with Mach number is consistent with the experimental results. 
Good agreement is observed between SUGAR-1D DSMC and experiment for a Mach 8 shock simulated with $\omega=0.75$.
Based on these values, the calculated ratio of shock-thickness to upstream bulk velocity, $L_s u_{x,1}^{-1}$, decreases with Mach number, as shown in figure~\ref{f:St}.
\vspace{\baselineskip}

The Strouhal number defined based on the above quantities and the weighted-averaged frequency, $f$, shown 
 in figure~\ref{f:FreqVsMach}, is relatively constant within a range of $St$=0.007 to 0.011 and a standard deviation  from 0.001 to 0.02.  This range of Strouhal numbers  is similar to those observed in the literature for highly compressible flows over embedded bodies.
For example, in the case of shock-dominated separated flows the Strouhal number associated with low-frequency shock motion ranges from 0.02 to 0.05~\citep[e.g.][]{dussauge2006unsteadiness,piponniau2009simple,clemens2014low,gaitonde2015progress,priebe2016low,tumuklu2018POF2} assuming that the characteristic length and velocity scales are given by the length of the separation bubble and the upstream bulk velocity, respectively. 
In the study of oblique SBLIs, \citet{nichols2017stability} defined the Strouhal number based on the upstream boundary-layer thickness and freestream velocity and found it to be within a range of 0.0003 to 0.05.
Therefore, we hypothesise that these fluctuations may also play a major role in shock-dominated flows having shock-thickness comparable to other important length scales in the flow, such as the size of the boundary layer in SBLIs and turbulent length scale in shock-turbulence interaction.  
\vspace{\baselineskip}

\begin{figure}[H]
\begin{center}
\includegraphics[width=0.50\columnwidth]{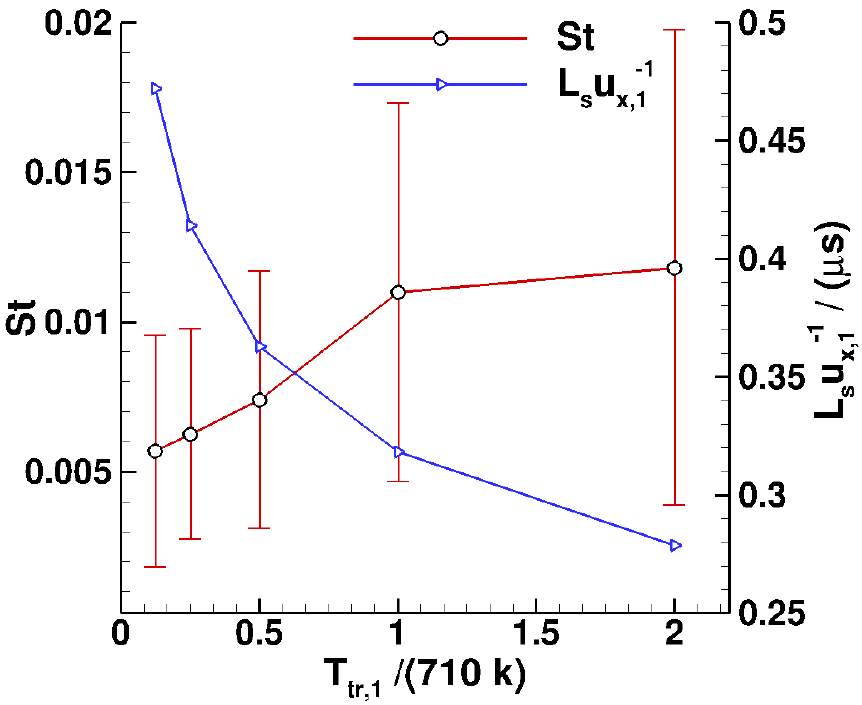}
\caption{Strouhal numbers for a range of cases simulated at Mach 7.2, as a function of  freestream temperature $T_{tr,1}$. A viscosity index of $\omega=0.81$ was used in the DSMC simulations and the inflection point in the NCE was found to be at 60\%.}
\label{f:StrouhalAtMach7dot2DiffTemp}
\end{center}
\end{figure}

To test whether the established range of Strouhal numbers holds for variations in upstream temperature, another set of DSMC cases were simulated in which the upstream temperature was changed from 710~K by fractions of 1/8, 1/4, 1/2, and 2, while the upstream bulk velocity was varied accordingly to maintain a constant Mach number of 7.2.
Figure~\ref{f:StrouhalAtMach7dot2DiffTemp} shows a decrease in the ratio of $L_su_{x,1}^{-1}$ with increase in temperature, due primarily to the increase in 
 $u_{x,1}$ rather than the shock-thickness.
The Strouhal numbers obtained using the aforementioned approach is within the range of St=0.005 to 0.011, with a small reduction with decrease in temperature.
Nonetheless, the $\pm$1 standard deviation of broadband frequencies is contained within the limits of $St$=0.001 to 0.02.

\section{Conclusion}~\label{sec:Conclusion}
The investigation of macroscopic fluctuations in the DSMC-computed Mach 7.2 shock layer revealed low and high frequencies on the order of tens and hundreds of kilohertz, respectively, in comparison to the freestream, which only exhibits high frequencies.
These disparities were attributed to the differences in particle distribution functions in the nonequilibrium zone of shock versus the equilibrium regime upstream.
The fluctuations in the normalized overall stress component, which is the mean of the PDF $f_{\xi_x}$, were found to be correlated with perturbations in the entire $X$-directional energy space of PDF.
Two distinct energy bins were identified depending on the sign of the cross-correlation coefficient, and a Lotka-Volterra type two-energy bin ODE model was constructed.
\vspace{\baselineskip}

The model accounted for the interaction of two energy bins through intermolecular collisions and particle fluxes from neighbouring computational cells.
At a location inside the shock, the model predicted two disparate time-scales: a longer time-scale (50-100~kHz) associated with the time of decay of fluctuations in the number of particles in energy bins to critical (or average) values and an order of magnitude smaller time-scale (281~kHz) associated with oscillations in the number of particles.
In the freestream, the model also predicted a decay time of fluctuations ten times longer than the mean-collision-time and, more importantly, an absence of low-frequency fluctuations consistent with the DSMC spectral analysis.
\vspace{\baselineskip}

Finally, a Strouhal number was defined based on the density gradient shock-thickness and upstream bulk velocity to nondimensionalize the low-frequency broadband characterised by a weighted average frequency and $\pm$1 standard deviation across a wide range of Mach numbers from 2 to 10.
The Strouhal number was found to range from $St$=0.007 to 0.011 with $\pm$1 standard deviation between 0.001 and 0.02, consistent with $St$ found in the literature on shock-dominated flows.
The established range of Strouhal numbers was also found to hold for a Mach 7.2 shock simulated with variations in upstream temperature.  
The presence of low-frequency fluctuations of shock layers suggests that these disturbances may play a key role in the receptivity process of transition in hypersonic flows, SBLIs, and shock-turbulence interactions, especially when the shock-thickness is comparable to other important length scales in the flow.

\newpage
\noindent{\bf  Supplementary data\bf{.}} \label{SupMat} Two supplementary movies are provided with the submitted manuscript.

\noindent{\bf Acknowledgements\bf{.}}
This work used the STAMPEDE2 supercomputing resources provided by the Extreme Science and Engineering Discovery Environment (XSEDE) at the Texas Advanced Computing Center (TACC) through allocation TG-PHY160006.
Additionally, S.S. would like to thank colleague Nakul Nuwal for helpful discussions regarding the two-energy bin model.
\\

\noindent{\bf Funding\bf{.}} The research conducted in this paper is supported by the Office of Naval Research under the grant No. N000141202195 titled, “Multi-scale modeling of unsteady shock-boundary layer hypersonic flow instabilities.”\\

\noindent{\bf 
Declaration of Interests\bf{.}} The authors report no conflict of interest. \\

\noindent{\bf  Author ORCID\bf{.}} Authors may include the ORCID identifers as follows.  S. Sawant, https://orcid.org/0000-0002-2931-9299; D. A. Levin, https://orcid.org/0000-0002-6109-283X; V. Theofilis, https://orcid.org/0000-0002-7720-3434.

\newpage
\appendix
\section{}\label{app:simplification}
This appendix describes the grouping of detailed collision processes $p_i$ into like-$P_i$ processes of table~\ref{tab:collProc} based on the energy density functions at locations $P$ and $F$.  
In addition we derive the simplifications to equation~\ref{dNA_sum_substituted} that allow us to reduce the six collision processes ($P_i$) given in table~\ref{tab:collProc} to four processes ($Q_i$).
\vspace{\baselineskip}

Starting with the reduction of collision processes $p_i$ to like-$P_i$ processes,  
Figure~\ref{f:binADist} shows that the energy distributions of $A_c$ and $A_d'$ are very similar at probes $P$ and $F$, so that $A_d'$-type particles can be considered to be the same as those of $A_c$.  
Additionally, although the distributions $f_{\xi_x}^A$ and $f_{\xi_y+\xi_z}^A$ of $A_c'$ and $A_d$ particles is the same at probe $F$, they are slightly different at probe $P$ at high $\xi_x$ and low $\xi_y+\xi_z$ energies.  
This occurs because when $A_c$ and $B$-type particles collide, the $B$ particle loses its $X$-directional energy which results in the increase of not only the $\xi_x$ of the $A_c$-type particle but also a noticeable increase in its $\xi_y+\xi_z$ energy, as can be seen in the transverse energy distributions of $A_c'$-type particles in figures~\ref{fb:zetaYZ_A_P} at probe $P$.
Also, when two $A_d$ particles collide, one of them is switched from energy bin $A$ to bin $B$ at the expense of higher transverse energy of the other $A_d$ particle that remains in bin $A$.
In the freestream, since these two $A_d$-type particles are the same $A_c'$-type particles that were generated from process $p_c$, their transverse energy distributions exactly match, as seen in figure~\ref{fd:zetaYZ_A_F}.
However, at probe $P$, it is also possible that only one of the two $A_d$ particles is an $A_c'$-type, and the other one has lower transverse and higher $X$-directional energy than the $A_c'$-type particle, as can be seen in the {\it slightly} larger $X$-directional energy distribution of $A_d$-type particles than the $A_c'$-type in figure~\ref{fa:zetaX_A_P} between $0.5 < \xi_x < 1.33$.  
We demonstrate that small distinctions such as this do not affect the dynamics of the system and assume that $A_c' \sim A_d$ to construct processes $P_c$ and $P_d$ given in table~\ref{tab:collProc}.  
\vspace{\baselineskip}

Similarly, we can describe the process $p_e$, where the particles of type-$A_{e}'$ have a very different transverse energy distributions shown in figures~\ref{fb:zetaYZ_A_P} than the other type-$A$ particles at probes $P$.
This process does not occur at probe $F$.
Ignoring the differences at low transverse energies, the distribution of $A_e'$ and $A_f$ particles is the same, as shown in figure~\ref{fb:zetaYZ_A_P}, therefore they can be denoted by the same identifier $A_e'$.
\vspace{\baselineskip}

\begin{figure}
    \centering
    \sidesubfloat[]{\label{fa:AssumtionForAa}{\includegraphics[width=0.48\textwidth, height=0.42\textwidth]{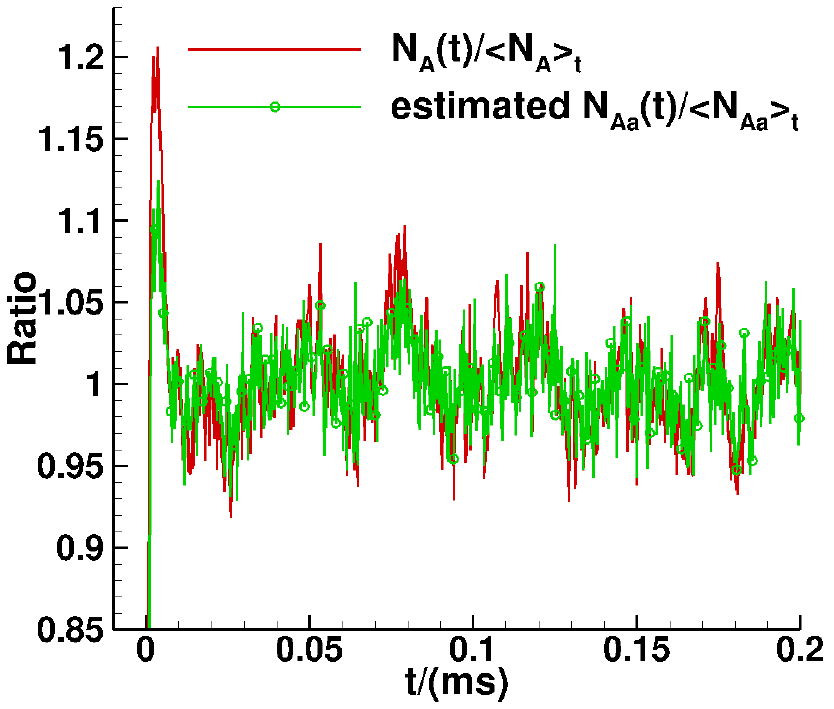}}}\hfill
    \sidesubfloat[]{\label{fb:AssumptionForAcprime}{\includegraphics[width=0.46\textwidth]{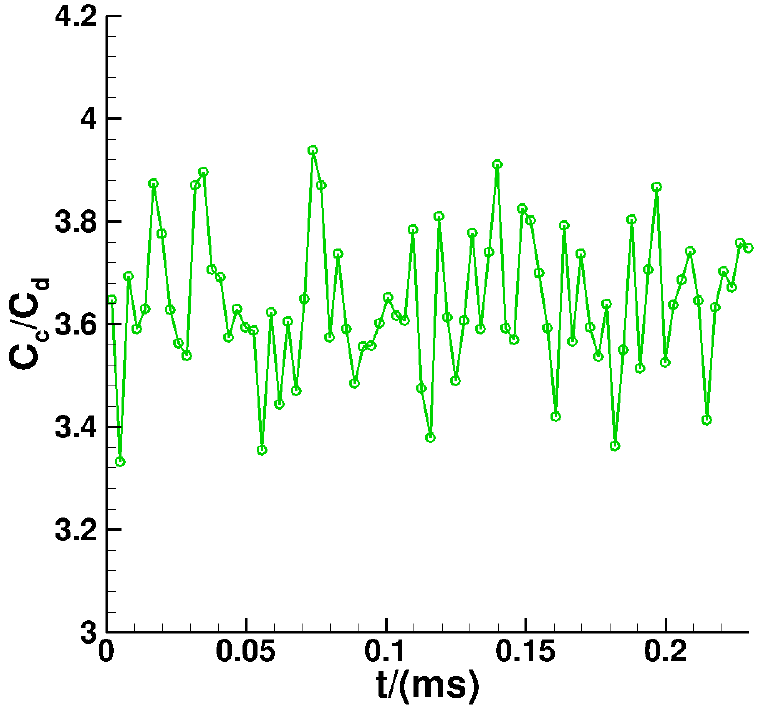}}}\,
    \caption{(\textit{a}) and (\textit{b}) show, at probe $P$, the comparison of time variation of ratio $N_{A_{a}}/\langle N_{A_{a}} \rangle_t$ with $N_{A}/\langle N_{A} \rangle_t$ and the time variation of ratio of collisions $C_c/C_d$, respectively, to justify assumptions in equations~\ref{Ratio_Na} and~\ref{Ratio_CcToCd}, respectively.
The data points in (\textit{a}) and (\textit{b}) are obtained by taking time-averages over a time window of 0.3~and 3~$\mu$s, respectively, to reduce statistical scatter.}
    \label{f:AssumptionVerify}
\end{figure}

To construct compound collision processes, $Q_i$ we start with 
 the rate coefficient $k_i$ for process $P_i$ which can be evaluated from the DSMC simulation as,
\begin{equation} 
\centering
k_{i} = \frac{\langle C_{i} \rangle_t}{\Delta t \langle N_{R_1} \rangle_t \langle N_{R_2} \rangle_t }
\label{RateCoeff_ki}
\end{equation}
where the collision takes place with between  particle types $R_1$ and $R_2$.
$\langle C_{i} \rangle_t$ for collision processes defined in table~\ref{tab:collProc} are listed in table~\ref{tab:DSMCNCols}.
Using the definition of rates $k_a$ and $k_b$, the first kinetic equation in equation~\ref{dNA_subspecies} can be written as,
\begin{equation} 
\centering
\begin{split}
\left[\frac{dN_{A_{a}}}{dt}\right]_{coll} &\approx - \frac{\langle C_a\rangle_t}{\Delta t \langle N_{A} \rangle_t \langle N_{B} \rangle_t} N_{A} N_{B} +\frac{\langle C_b\rangle_t}{\Delta t \langle N_{{B}}\rangle_t^2 }  N_{B}^2 \\
\left[\frac{dN_{A_{a}}}{dt}\right]_{coll} &\approx -\tilde{k}_a N_{A} N_{B} + k_b N_{B}^2 \\
\end{split}
\label{Modified1stKinetic}
\end{equation}
where we have used the fact that,
\begin{equation} 
 \frac{N_{A_{a}}}{\langle N_{A_{a}} \rangle_t} \approx \frac{N_{A}}{ \langle N_{A}\rangle_t }
\label{Ratio_Na}
\end{equation}
as shown in figure~\ref{fa:AssumtionForAa}.
The ratio $N_{A_{a}}/\langle N_{A_{a}} \rangle_t$ is estimated from figure~\ref{fa:zetaX_A_P} by obtaining the number of particles between $0.0 < \xi_x < 1.33$ at probe $P$.
At probe $F$, the ratio of $N_{A_a}/\langle N_{A}\rangle_t$ cannot be directly estimated from figure~\ref{fc:zetaX_A_F} because the distribution of $A_a$-type particles overlaps with other $A$-type particles.
Yet, equation~\ref{Ratio_Na} is assumed to hold true under the assumption that the entire $X$-directional energy zone $-28<\xi_x<0$ is coarsened into a single bin $A$.
\vspace{\baselineskip}

Using the definitions of rates $k_c$ and $k_d$, the second kinetic equation in equation~\ref{dNA_subspecies} can be written as,
\begin{equation} 
\left[\frac{dN_{A_{c}}}{dt}\right]_{coll} = - \frac{\langle C_c \rangle_t}{\Delta t \langle N_{A_{c}} \rangle_t \langle N_{B} \rangle_t} N_{A_{c}} N_{B} +\frac{\langle C_{d} \rangle_t}{\Delta t \langle N_{A_{c}'}\rangle_t^2 }  N_{A_{c}'}^2 \\
\label{dNAcdt_expanded}
\end{equation}
Since the average number of $A_{c}'$ particles are not known, it is difficult to estimate the second term on the right hand side.
However, this term can be simplified by observing from figure~\ref{fb:AssumptionForAcprime} that  the ratio of rates $C_c/C_d$ does not vary significantly from the average rates at probe $P$, i.e.,
\begin{equation} 
\centering
\begin{split}
\frac{C_c}{C_d} \approx \frac{\langle C_c \rangle_t}{\langle C_d \rangle_t}
\end{split}
\label{Ratio_CcToCd}
\end{equation}
and the standard deviation in the fluctuation of the ratio of $C_c/C_d$ is only 3.7\% of the average ratio of 3.63 for probe $P$ and 4.2\% of an average ratio of one for probe $F$. Using equation~\ref{RateCoeff_ki} for $C_{c}$ and $C_{d}$ in \ref{Ratio_CcToCd}, and the definition of $k_c$ based on the average rate, we obtain,
\begin{equation} 
k_d N_{A_{c}'}^2 \approx  \frac{\langle C_d \rangle_t }{\Delta t \langle N_{A_{c}} \rangle_t \langle N_{B} \rangle_t} N_{A_c} N_B \\
\label{kd_NacPrimeTerm}
\end{equation}
In addition, we can see from figures~\ref{fa:zetaX_A_P} and~\ref{fb:zetaYZ_A_P} that the $A_c$ particles have the same energy distributions as the $A_h$ at probe $P$, which implies that any particle that takes part in process $P_c$ can take part in process $P_h$ and together they constitute 72.37\% of total collisions in which the particles in energy bin $A$ take part.  
Therefore, we can expect the changes in the ratio $N_{A_c}/\langle N_{A_c} \rangle_t$ are correlated with the changes in ratio $N_A/\langle N_{A}\rangle_t$ and write,
\begin{equation} 
\centering
\begin{split}
\left[\frac{dN_{A_{c}}}{dt}\right]_{coll} &\approx - \frac{\langle C_c \rangle_t - \langle C_d \rangle_t}{\Delta t \langle N_{A} \rangle_t \langle N_{B} \rangle_t} N_{A} N_{B} \\
                                          &\approx -\tilde{k}_cN_A N_B \\
\end{split}
\label{Modified2ndKinetic}
\end{equation}
The same assumption holds at probe $F$ because the $A_c$-type particles constitute the majority of particles from $-20<\xi_x<-5$, which is a significant portion of energy bin $A$ ($-28<\xi_x<0$), as seen from figure~\ref{fc:zetaX_A_F}.
\vspace{\baselineskip}

By using the aforementioned assumptions, the third kinetic equation in equation~\ref{dNA_subspecies} becomes,
\begin{equation} 
\centering
\begin{split}
\left[\frac{dN_{A_{c}'}}{dt}\right]_{coll} &= 2 k_c N_{A_{c}} N_{B} - 2 k_d N_{A_{c}'}^2 \\
                                           &\approx 2 \tilde{k}_c N_{A} N_{B} 
\end{split}
\label{Modified3rdKinetic}
\end{equation}
We also drop the second term on the right hand side of the fourth kinetic equation in equation~\ref{dNA_subspecies} because $\langle C_e \rangle_t  >> \langle C_f \rangle_t$.  
Note that both $\langle C_e \rangle_t = \langle C_f \rangle_t = 0$ at probe $F$.
\vspace{\baselineskip}
 
By substituting equations~\ref{Modified1stKinetic},~\ref{Modified2ndKinetic}, and ~\ref{Modified3rdKinetic},  into equation~\ref{dNA_sum_substituted}, we obtain,
\begin{equation} 
\centering
\begin{split}
\left[\frac{dN_A}{dt}\right]_{coll} &= -\tilde{k}_{a} N_{A} N_{B} + k_b N_{B}^2 + \tilde{k}_{c} N_{A} N_{B} + 2 k_e N_{B}^2 \\
\end{split} 
\label{dNA_sum_simplified}
\end{equation}
which when substituted into equations~\ref{ODEModel} and~\ref{dNA_dt_equal_dNB_dt} gives the final system of dynamic equations, equation~\ref{ODEModel_Final}.

\newpage
\bibliographystyle{jfm}
\bibliography{References}

\end{document}